# Metalated Porous-Organic-Polymer Renders Mustard-Gas Simulant Harmless: Core Planarity Matters

Ratul Paul,[a, b],‡ Chitra Sarkar, [a, b], ‡ Manjari Jain,[c], ‡ Shaojun Xu, [d, e] Kashmiri Borah,[f] Duy Quang Dao,[g] Chih-Wen Pao,[h] Saswata Bhattacharya*,[c] & John Mondal*,[a, b]

[a] R. Paul, C. Sarkar and Dr. J. Mondal
Catalysis & Fine Chemicals Division, CSIR-Indian Institute of Chemical Technology, Uppal Road, Hyderabad 500 007, India. johnmondal@iict.res.in;
Email: johnmondal@iict.res.in, johncuchem@gmail.com (J.M.)

[b] R. Paul, C. Sarkar and Dr. J. Mondal
Academy of Scientific and Innovative Research (AcSIR), Ghaziabad- 201002, India.

[c] M. Jain and Dr. S. Bhattacharya
Department of Physics, Indian Institute of Technology Delhi, Hauz Khas, New Delhi 110 016, India.
Email: saswata@physics.iitd.ac.in  (S.B.)

[d] Dr. S. Xu
Cardiff Catalysis Institute, School of Chemistry, Cardiff University, Cardiff CF10 3AT, U.K.

[e] Dr. S. Xu
UK Catalysis Hub, Research Complex at Harwell, Rutherford Appleton Laboratory, Harwell, OX11 0FA, United Kingdom.

[f] K. Borah
Polymers & Functional Materials Division, CSIR-Indian Institute of Chemical Technology, Uppal Road, Hyderabad 500 007, India.

[g] Dr. D. Q. Dao
Institute of research and development, Duy Tan University, Da Nang, 550000, Viet Nam.

[h] Dr. C. W. Pao
National Synchrotron Radiation Research Center, 101 Hsin-Ann Road, Hsinchu 30076, Taiwan.

‡ R.P., C.S., and M.J. are equally contributed to this work.

Supporting information for this article is given via a link at the end of the document.

**Abstract:** The presence of open active metal sites in Metal-Organic Frameworks (MOFs) exhibit higher catalytic activity. However, rational accomplishment of MOFs in heterogeneous catalysis is limited due to coordination bonds. Recently balanced characteristic feature with combination of both the covalent bonds (structural stability) and open metal sites (single site catalysis) introduced an entirely organic alternative architecture named as Metalated Porous-Organic-Polymers (M-POPs). In this contribution, we demonstrate successful construction of two Fe-POPs (**Fe-Tt-POP** & **Fe-Rb-POP**) by ternary copolymerization approach for catalytic oxidative decontamination of different sulfur-based mustard gas simulants. **Fe-Tt-POP** exhibited superior catalytic performance for oxidation of the thioanisole (TA) studied in terms of conversion (99% after 13 h) in comparison with **Fe-Rb-POP** (43% after 13h). The remarkable difference in the mechanistic pathways towards catalytic performance for oxidation of TA was investigated by *in situ* operando Diffuse Reflectance Infrared Fourier Transform Spectroscopy (DRIFTS) analysis, complemented by Density Functional Theory (DFT) computational study.

## Introduction

Metal-Organic Frameworks (MOFs) represent an emerging class of Advanced Porous Materials (APMs) which mainly comprising with an extensive combination of metal ions and organic linkers thereby offering a diverse range of functionalities & porous structures. But the realistic implementation of these porous materials in robust heterogeneous catalysis has been restricted owing to their nature & strength of inherent coordination bond involved.[1] Although in very recent times, Porous-Organic-Polymers (POPs) have also garnered colossal research interests because of their interesting characteristic features such as high mechanical and chemical stabilities, adjustable chemical functionality with the modification by varying synthetic approaches and selection of organic building blocks (monomers), specific surface area and well-defined pores.[2-6] In contrast, instead of presence of highly cross-linked rigid covalent bonds, the metal-free POPs become unable to accomplish various sophisticated chemical and physical properties with advanced applications leading to the development of modern science.[7-9] In this regard to address these associated drawbacks, Metalated Porous-Organic-Polymer (M-POPs) would be ideal substitute which could be indisputably developed with the combination of polymer chemistry & inorganic chemistry. These M-POPs can create the bridge in between MOFs & POPs thereby attributing to the self-complementary balanced characteristic features which comprises both the covalent bonds (structural stability) & open metal sites (single-site catalysis).[10] Generally, various synthetic approaches have been reported to construct M-POPs which includes (a) direct synthesis; (b) subcomponent self-assembly & (c) post-synthetic metalation, respectively using metal-coordinating molecules or complexes as building blocks. Although, direct synthetic strategy is the most straightforward and smart approach among them where the utilization of prevalent metal precursors can effectively expand versatile functionalities of POPs but some limitations including sensitivity, stability, solubility, and compatibility are associated with it.[11]

In 1917, sulfur mustard, also known as mustard gas (HD) was first generated in large scale to utilize as Chemical Warfare Agent (CWA) during World War I, holding the title "weapon of choice".[12-15] Including severe blistering over the exposed skin, HD has also the ability to effect on eyes and respiratory system, even cause death on higher doses.[16] Recently, international ban over its utilization, storage and production are unable to restrain its popularity as CWA.[16] Bleach powders with the varying formulations were the first used as effective decontaminants for the HD, but they suffered from various drawbacks including (i) significant diminishment of the activity with time; (ii) huge quantity of bleach is employed for decontamination & (iii) its corrosive nature. Very recently, significant advancement has been promoted to design Metal Oxides,[17] Polyoxometalates, [18,19] MOFs, [20-22] Hydrogen-bonded Organic Framework (HOF),[23] & Polymer [24] for the decontamination of the HD, but their practical implementation on battlefield for real-time protection is limited. Till now several oxidative routes including hydrolysis, dehydrohalogenation & oxidation of HD detoxification are explored, but the presence of easily oxidizable bivalent sulfur is the main attractiveness of the oxidation step.

Considering exceptional advantageous characteristic features of M-POPs, we have developed an emerging functional material (Fe-POPs) by ternary copolymerization strategy. In this study, the as-synthesized **Fe-POPs** demonstrates superior catalysis performance for catalytic decontamination of HD. In addition, we have also shown that the catalytic binding pocket inside the porous network, alignment & proper orientation of the POPs control the product selectivity. Furthermore, to gain in depth understanding about the mechanistic pathway involved, the role of catalytic binding pocket, and the core planarity in the **Fe-POP** based catalytic systems *in-situ* operando Diffuse Reflectance Infrared Fourier Transform Spectroscopy (DRIFTS) spectroscopy study & Density Functional Theory (DFT) [25,26] investigations were performed.

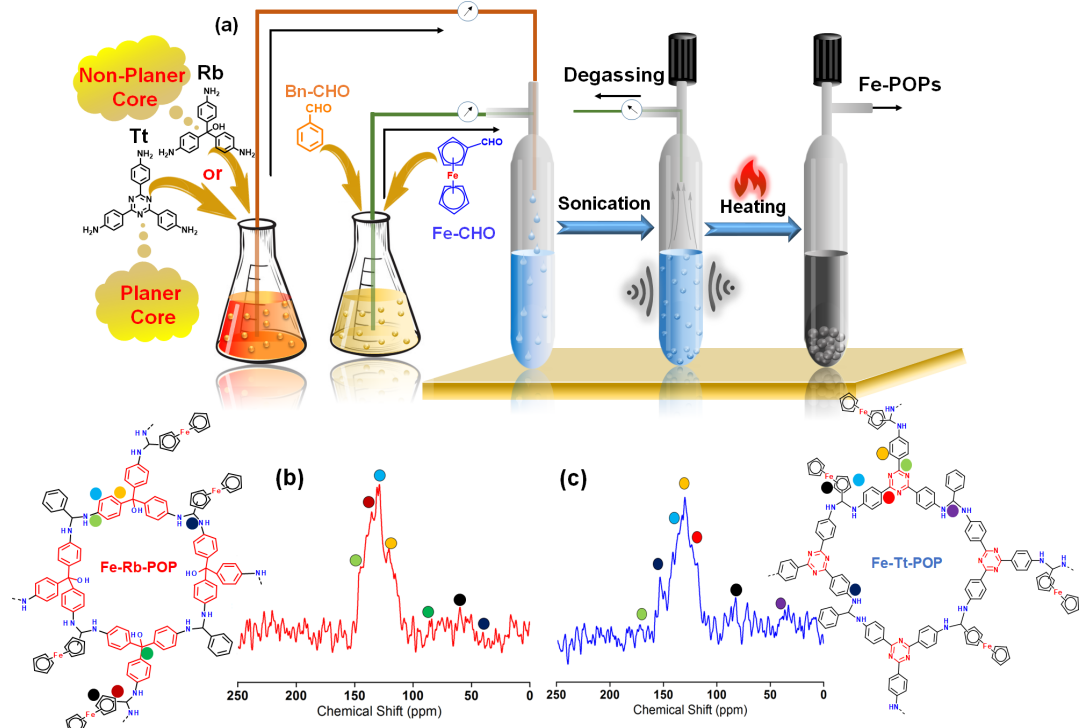

**Figure 1:** (a) Schematic Illustration Showing the Synthetic Strategy of as-synthesized **Fe-Rb-POP** & **Fe-Tt-POP** respectively; $^{13}$C-CP solid state MAS NMR spectra of **Fe-Rb-POP** (b) & **Fe-Tt-POP** (c), respectively.

## Results and Discussion

We introduced ternary copolymerization strategy *via* facile and economic Schiff base reaction with the ferrocenealdehyde (**Fe-CHO**), benzaldehyde (**Bn-CHO**), 4,4',4''-(1,3,5-Triazine-2,4,6-triyl) trianiline (**Tt**), and Pararosaniline Base (**Rb**) to furnish two terpolymers (**Fe-Tt-POP** & **Fe-Rb-POP**), respectively, with the aminal-linkages (-NH-CH-NH-). We have initiated the reaction under sealed-tube condition in Dimethyl sulfoxide (DMSO) solvent at 140º C temperature (Figure 1a). The detailed procedure was described in the experimental section (see SI). Among all the conditions screened, the combination of Fe-aldehyde, aromatic amine & aromatic aldehyde (1:2:1) and DMSO as a solvent was the best condition for the synthesis of Fe-metalated POP with high yield (~89%). Both **Fe-Tt-POP** & **Fe-Rb-POP** are black powders, and they are insoluble in water and common organic solvents. N, C and H contents in elemental analysis (EA) was detected for the both **Fe-POPs**, presented in Table S1, SI. Thermogravimetric analysis of our as-synthesized **Fe-POPs** have been performed to unveil the thermal stability of the porous organic polymer, shown in Figure S2, SI.[27] In case of **Fe-Rb-POP**, almost 20% weight loss was experienced in the range of 200 to 300º C, demonstrating the fragmentation of hydroxyl functional groups of polymeric networks in the form of small molecules such as $CO_2$ or $H_2$. Interestingly, overall, 30% weight loss up to 600º C clearly confirms the convenient thermal stability of **Fe-Rb-POP**. However, **Fe-Tt-POP** exhibited superior thermal stability compared to **Fe-Rb-POP** with gradual drop of ~20 % weight from 350º C to 420º C due to the functional group decomposition to the small molecules. No further prominent weight loss with continuous heating illustrates superb thermal as well as mechanical stability of our **Fe-Rb-POP** catalyst. Inductively Coupled Plasma Mass Spectrometry (ICP-MS) analysis indicated that the Fe contents in **Fe-Tt-POP** & **Fe-Rb-POP** are 2.0 & 2.1 wt%, respectively. Further, the as-synthesized **Fe-POP** catalysts were comprehensively characterized *via* different techniques like Solid-state Cross-Polarization Magic Angle Spinning Carbon-13 Nuclear Magnetic Resonance ($^{13}$C-CP MAS NMR), wide-angle Powder X-ray Diffraction pattern (PXRD), Fourier Transform Infrared Spectroscopy (FT-IR), $N_2$-Physisorption, Transmission Electron Microscopy (TEM), Field Emission Scanning Electron Microscopy (FE-SEM), X-ay Photoelectron Spectroscopy (XPS), X-ray Absorption Spectroscopy (XAS) analysis. $^{13}$C-CP MAS NMR spectra revealed molecular connectivity and the chemical environment of the carbon nuclei of the ferrocene based porous organic polymeric network. **Fe-Rb-POP** exhibited four resonance signals at *δ* = 145, 128, 120 & 82 ppm, respectively, which could be attributed to the carbon atom adjacent to the N, aromatic carbons of benzene ring as well as aliphatic carbon attached to the hydroxyl group of the pararosaniline moiety (Figure 1b).[28] Presence of core triazine ring was validated by the signal at *δ* = 171 ppm, whereas three strong signals emerging at *δ* = 129, 122 & 153 ppm, respectively, corresponding to the different chemical environment around the carbon atoms of the benzene ring for the **Fe-Tt-POP** (Figure 1c).[29] All the carbon atoms in the $^{13}$C-CP MAS NMR spectra are marked to define the structural integrity of the polymer backbone units. Successful copolymerization to produce these terpolymers could also be evidenced by the chemical shifts at $\delta$ = 40 & 36 ppm, respectively, which unambiguously signifies the tertiary carbon atom of (-NH-CH-NH-) aminal-linkages.[30] Two resonance signals appeared at $\delta$ = 70, 134 ppm, could be assigned to the aromatic carbons of cyclopentadienyl (Cp) ring.[30,31] FT-IR was performed to validate the 3D polyaminal network of the porous organic polymer (Figure 3a). It was already established that under suitable reaction condition imine double bond normally formed upon condensation of aldehyde & amine, followed by the further addition of primary amine group, generated this aminal-linkage.[30] The characteristic stretching bands at 3237 cm$^{-1}$ & 3105 cm$^{-1}$ could be assigned as the secondary amine (N-H) & newly formed methylene (-CH-) for **Fe-Rb-POP** framework (Figure 3a).[32] Furthermore, distinct vibrations at 1122 cm$^{-1}$ & 914 cm$^{-1}$, indicate the incorporation of ferrocene moiety to the polymeric framework.[30,31] A broad vibration at 3524 cm$^{-1}$, could be attributed to the hydroxyl functional group of the pararosaniline moiety. The stretching vibrations appeared at 1372 cm$^{-1}$ confirmed the presence of triazine ring for **Fe-Tt-POP** (Figure 3a).[33] The stretching frequency of secondary amine (N-H) & newly formed methylene (-CH-) linkages were detected at 3252 cm$^{-1}$ & 3120 cm$^{-1}$, respectively.[27] In addition, stretching vibration bands located at 1110 cm$^{-1}$ & 930 cm$^{-1}$ confirmed the presence of ferrocene in the **Fe-Tt-POP** polymer matrix.[33] The above-mentioned results manifested the successful integration of all monomer units to polymeric network through aminal-linkage. High stability, unique rigid and double-deck sandwich-like structure of ferrocene monomers make them special in the design of metalated porous network among other various metal-containing monomers.[34] From the structural point of view, one of the crucial factors engender an appropriate catalytic pocket with possible local structure, in which convergent binding sites are positioned in such a way that substrate molecules can be held in close proximity. Moreover, the core planarity along with the alignment with proper orientation in the **Fe-Tt-POP** compared with the **Fe-Rb-POP** also reinforces the catalytic activity and product selectivity.[35]

The structural and electronic properties of **Fe-POPs** are shown in the Figure 2a to 2f. The optimized geometry of the POP display a pore size varied from about 10 to 14 Å (Figure S1a, SI). The Electrostatic Potential (ESP) maps show that the most negative charges are found at the tris(4-aminophenyl)-methanol moieties whereas the most positive charges are found at amine groups of the compound (Figure 2c). Moreover, the Highest Occupied Molecular Orbital (HOMO) is locally distributed at some aniline groups (Figure 2a) whereas the Lowest unoccupied Molecular Orbital (LUMO) distributions are only found at the ferrocenyl moieties (Figure 2b). This observation suggests the reactive zones of **Fe-Rb-POP** compound in the reactions with the external adsorbate compounds. While Figure 2d to 2e represents the optimized geometry, the HOMO and the LUMO distributions as well as the ESP maps calculated in the gas phase for the **Fe-Tt-POP** structure. The geometrical properties calculation shows that the pore size of one POP unit is about 15 Å with a square shape (Figure S1b, SI). Both the HOMO and LUMO distribute along the four edges of the POP unit (Figure 2d & 2e). It is worth noting that the frontier orbitals are not found at the ferrocenyl moieties. Thus, it can deduce that the 2,4,6-tris(4-aminophenyl)- 1,3,5-triazine moieties, where the electron density transfer processes with the adsorbates could occur, play an important role in the reactivity of the POP material. In addition, the ESP maps calculation (Figure 2f) shows that the most negative charged zones with the red color are found at the aromatic rings distributed along the four edges of the pore, whereas the four corners display the most positive charges with blue color.

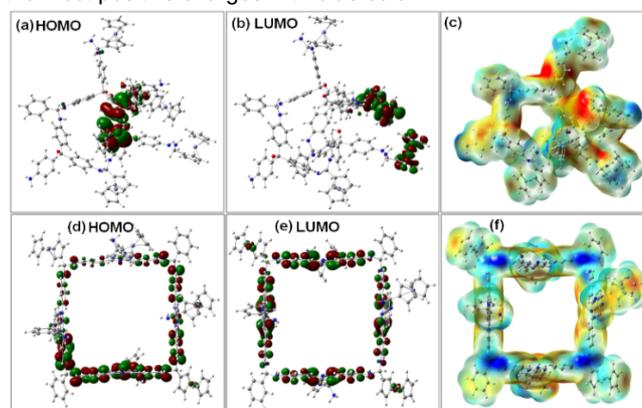

**Figure 2:** (a & d) HOMO; (b & e) LUMO, and (c & f) ESP maps of **Fe-Rb-POP** and **Fe-Tt-POP**, respectively. All calculations are performed in the gas phase at the B3LYP/Def2-SV(P) level of theory. Iso-value of ESP map is 0.0004. Iso-value of the orbitals is 0.02.

The wide-angle X-ray powder diffraction patterns of the respective **Fe-Rb-POP** & **Fe-Tt-POP** (Figure S3, SI) exhibit broad peak in the region of Bragg's angle 15° to 28° which clearly indicates the amorphous nature of two ferrocene-based POPs.[36] To evaluate the permanent porosity of the Fe-porous organic polymer, N$_2$-adsorption/desorption isotherms analysis at 77 K was carried out (Figure S4, SI). Both the material comprises typical type IV isotherm with gradual N$_2$ uptake and a large H3-type hysteresis loop at high $P/P_0$ (0.4-0.8) region which is characteristic feature of the mesoporous materials. Generally large hysteresis loop is reflected in the POP based materials owing to their flexible elastic expansion of restricted access pores (nominally closed pores) and swelling of polymeric framework leading to irreversible gas uptake during gas adsorption.[37] The Brunauer-Emmett-Teller (BET) surface area & total pore volume of two Fe-based POPs (**Fe-Rb-POP** & **Fe-Tt-POP**) were calculated as (245 m$^2$g$^{-1}$, 0.232 cm$^3$g$^{-1}$) & (296 m$^2$g$^{-1}$, 0.238 cm$^3$g$^{-1}$), respectively. The Pore-Size Distributions (PSD) of the polymeric materials (Figure S5, SI) as measured by using Non-Local Density Functional Theory (NLDFT) demonstrated the narrow pore size distributions with the pore predominately arising at 3.5-3.9 nm, corresponds to mesoporous nature.[38] Surface morphology as well as microstructure of the as-synthesized **Fe-Rb-POP** & **Fe-Tt-POP** hybrid materials has been illustrated by FE-SEM analysis. We observed a parched-earth type of morphology with the appearance of chunks for the rosaline based porous organic polymer, which is due to the random agglomeration of particles during high temperature synthesis process (Figure S6, SI). In case of **Fe-Tt-POP**, FE-SEM microscopic image reveals the similarity with the satellite image of a hill station, probably accumulation of Fe based porous organic polymer at high temperature synthesis process could be the reason behind this unique structure with highly rough surface. On the further inspection of surface morphology through TEM analysis, the surface roughness is clearly recognizable for both catalysts (**Fe-Rb-POP** & **Fe-Tt-POP**), displayed on Figure S7, SI. Highly magnified TEM images demonstrate spider web type of morphology, where the darker & gloomy area represents the emergence metal & polymeric network individually (Figure 3b & 3d). The probable reason behind this specific spider web structure is hyper cross linking between monomers as well as a very asymmetrical distribution of metal particles over the polymeric framework. We can also clearly observe almost homogeneous distribution of metal precursor over polymeric network for the two catalyst **Fe-Rb-POP** & **Fe-Tt-POP**. The High-angle Annular Dark-field Scanning Transmission Electron Microscopy (HAADF-STEM) image of the selected areas with the corresponding elemental mapping analysis demonstrated (Figure 3c & 3e) Fe (violet), C (green), N (red) & O (blue) are nearly distributed in the same location accompanying with each other.[4]

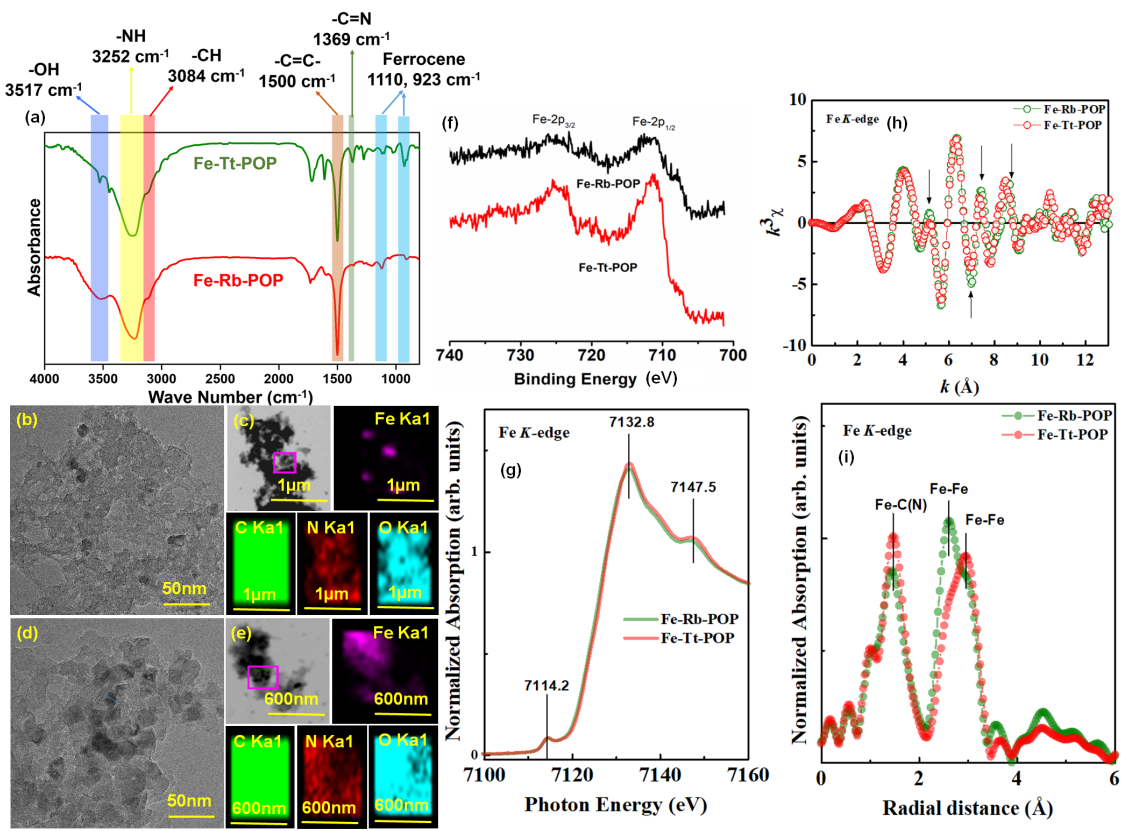

**Figure 3:** (a) FT-IR spectra of **Fe-Tt-POP** and **Fe-Rb-POP**. TEM image of **Fe-Rb-POP** (b, c) and **Fe-Tt-POP** (d, e) along with corresponding elemental mapping of Fe (violet), C (green), N (red) & O (blue) images are provided. Core-level XP spectra of Fe-2p region (f) for **Fe-Rb-POP** & **Fe-Tt-POP** materials, respectively. (g) Fe $K$-edge XANES spectra, (h) the $k^3$-weighted EXAFS oscillations spectra, (i) Fourier transform magnitudes of the Fe $K$-edge EXAFS showing local atomic distribution around Fe in **Fe-Rb-POP** & **Fe-Tt-POP**, respectively.

The X-ray Photoelectron Spectroscopy (XPS) is an authentic quantitative spectroscopic technique, which illustrates elemental composition as well as chemical & electronic state of each element on the surface of a material by photoelectric effect. The survey spectra provided on Figure S8, clearly indicates the presence of C, N, O & Fe characteristic peaks for both **Fe-Rb-POP** & **Fe-Tt-POP**, respectively. High resolution C-1s XPS spectra (Figure S9, SI) can be deconvoluted into three different types of binding energies, which could be attributed to the (284.4 eV) C-C/C=C, (285.6 eV) C-OH & (290.4 eV) C-N/C=N bonding in the POP frameworks. Deconvoluted N-1s XPS spectrum of **Fe-Rb-POP** (Figure S10, SI) appeared at 399.8 eV & 401.0 eV could be assigned to C-N & N-H type of bonds. In the deconvoluted N-1s XPS spectrum of **Fe-Tt-POP** (Figure S10, SI), three distinguishable binding energy peaks located at 398.0 eV, 399.8 eV & 401.0 eV, corresponds to triazine unit, C-N & N-H, respectively. High resolution XPS spectra of Fe-2p (Figure 3f) exhibits characteristic peaks at 711.5 eV, 711.4 eV & 725.2 eV, 724.7 eV for **Fe-Rb-POP** & **Fe-Tt-POP**, respectively. The peaks appearing at ~ 711 eV & ~ 725 eV corresponds to the Fe-2p$_{1/2}$ & Fe-2p$_{3/2}$, respectively, signifying existence of Fe (II) in both POP systems.[39,40] X-ray Absorption Spectroscopy (XAS) is an exclusive spectroscopic technique to unveil the electronic and structural details of a particular catalyst during the reaction in presence of reactant. Recently, XAS technique is able to provide geometric information of catalytic systems with short-range order through conventional Extended X-ray Absorption Fine Structure (EXAFS) technique and discover the adsorbate coverage as well as binding site information via X-ray Absorption Near-Edge Structure (XANES). Herein, difference in catalytic activity of **Fe-Rb-POP** and **Fe-Tt-POP** catalyst, the establishment of local geometric/electronic structure through X-ray Absorption Fine Structure Spectroscopy (XAFS) is utterly important. Fe $K$-edge X-ray Absorption Near Edge Structure (XANES) of both catalysts exhibited similar kind of spectral feature with a pre-edge peak at ~7114 eV (Figure 3g) for mixed metal Fe$^{+3}$/Fe$^{+2}$. The pre-edge peak appeared due to the dipole forbidden 1s→3d transition for the centrosymmetric molecule and its position is eventually modified the iron valance. Zarbin *et al.* established the fact that α-Fe$_2$O$_3$ standard sample and ferrocene based PVG/Fc-air-2 h sample exhibited the pre-edge peak at same energy whereas in case of neat ferrocene the peak lowered to 2 eV.[41] This result indicates the increase in oxidation state of Fe in PVG/Fc-air-2 h sample, corresponds to ferricenium occurrence[41] intensity XANES signal appeared at ~7132 eV than that of signal at ~7147 eV could be attributed to the translocation of the central Fe atoms away from the in plane of POP framework (Figure 3g). The k space EXAFS oscillation curves (Figure 3h) of both the samples emerged quite similar to each other but some variations in the oscillation shape and amplitude are observed. This slight difference in the peak intensities arising at ~5.1, ~7.0, ~7.43 & 8.74 Å (marked with arrow) also indicates the modification of local atomic environment of Fe in the two different POPs.[42] The Radial Distribution Function (RDF) from the Fourier-transformed EXAFS spectra (Figure 3i) showed existence of a strong peak at the R space of 1.4 Å may be attributable to the Fe-N(C) shell, whereas the decrease in intensity for **Fe-Rb-POP** indicates the lowering in coordinating N(C) atoms to the Fe center.[43] Surprisingly, the difference in the environment of two-catalysts confirmed by the lowering in R space value from 2.9 to 2.6 Å for Fe-Fe shell in case of **Fe-Rb-POP**. Similar trend was observed in case of MOF derived Cu/CuFe$_2$O$_4$ catalyst due to the change in Fe environment with Cu atom.[44] This result clearly coincides with the diverse feature achieved from TEM analysis. Fourier transform Fe *K*-edge EXAFS also reveals Fe-C intramolecular bond length appeared to 1.92 Å, which is very similar to the Fe-C bond belongs to Cp (cyclopentadiene) unit.[45] The Debye-Waller factors ($\sigma^2$) are evaluated to be 0.01058 & 0.0119 from the fitting EXAFS results of **Fe-Tt-POP** & **Fe-Rb-POP**, respectively. A slight increase in $\sigma^2$ value for **Fe-Rb-POP** refers to the increase in the framework structural asymmetry with disorder in local atomic environment. In summary, we can conclude that the diverse geometry obtained from different monomeric unit will definitely reflected in varying catalytic activity.

## Catalytic Performance of Fe-POPs

We have executed decontamination of sulfur mustards with the as-synthesized two novel porous organic polymers **Fe-Tt-POP** & **Fe-Rb-POP** with the utilization of the local catalytic pocket environment inside the porous network. In this study, thioanisole (TA) sulfur ether was employed as a model compound for the partial catalytic oxidation of sulfide (Ph-SCH$_3$) to sulfoxide (Ph-SOCH$_3$) (MPS), without forming over oxidized sulfone (PhSO$_2$CH$_3$) (MPSF) derivative which is more toxic and harmful compared with the other counterpart. In this investigation, our initial experiment began with the catalytic oxidation of TA (0.1 ml, 0.5 mmol) with H$_2$O$_2$ (0.25 ml) as an oxidant over the two **Fe-POP** catalysts in acetonitrile at 100°C oil-bath temperature. The progress of the catalytic reaction with the distribution of reactants and product components in the resulting mixture after the completion of the reaction is determined by the GC-FID technique based on an authentic sample. Evolution of reactant and product distributions against time (h) of both the catalysts has been displayed in Figure 4. As shown in Figure 4c, **Fe-Tt-POP** demonstrated the most superior catalytic performance for oxidation of the TA studied in terms of conversion (99% after 13 h). In this regard, we have achieved 94% selectivity towards sulfoxide (MPS) with 5% sulfone (MPSF) for **Fe-Tt-POP** in comparison with the 48% conversion after 13 h with 43% selectivity toward sulfoxide (MPS) for **Fe-Rb-POP** (Figure 4b). Notably, this compelling amount of difference in catalytic performance perhaps refer to the proper alignment & orientation of catalytic pocket inside the porous framework in the case of planar structure of **Fe-Tt-POP**. This also offers favorable binding interaction with the close proximity arrangement of substrates inside the catalytic core.[46]

Similar experiments were carried out under different reaction conditions like without catalyst, without H$_2$O$_2$, under N$_2$ atmosphere, and at room temperature (Table S3, SI) with the best performing **Fe-Tt-POP** catalyst. All the above-mentioned experiments exhibit a very less amount of TA conversion with a trace amount of sulfoxide (MPS) production. We have also performed the catalytic oxidation under various temperatures (Figure 4d & 4e) keeping the other reaction parameters constant for the comparison study. At 60º C, **Fe-Tt-POP** catalyst shows 35% TA conversion with 34% sulfoxide (MPS) selectivity, while **Fe-Rb-POP** shows only 17% conversion with 15% selectivity.

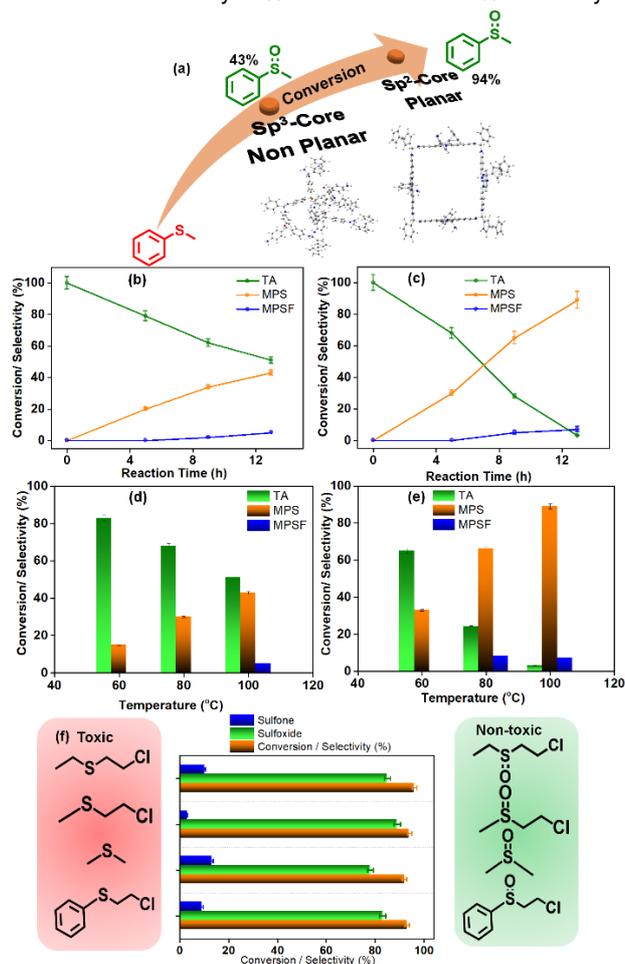

**Figure 4:** Distribution of reactant and product against time for the catalytic oxidative transformation of thioanisole (TA) to methyl phenyl sulfoxide (MPS) and methyl phenyl sulfone (MPSF) by using **Fe-Rb-POP** (a) & **Fe-Tt-POP** (b). Influence of reaction temperature (º C) for catalytic oxidation of TA with **Fe-Rb-POP** (c) & **Fe-Tt-POP** (d), respectively. Catalytic Oxidative decontamination of different sulfur-based mustards gas simulants using **Fe-Tt-POP** under similar reaction conditions (e). Reaction Conditions: TA (0.5 mmol), H$_2$O$_2$ (0.25 ml) CH$_3$CN (10 ml), catalyst (50 mg), temperature (100 °C), and time (13 h).

With increasing temperature up to 80º C, both the conversion (35 to 76%) and selectivity (33 to 65%) increases for **Fe-Tt-POP**. At 100º C, **Fe-Tt-POP** exhibits maximum catalytic activity with 100% TA conversion, while only 48% conversion was observed in the case of **Fe-Rb-POP**. Further examination on the catalytic property with various Fe-based catalysts including homogeneous counterparts obviously revealed the superior performance of our Fe-based terpolymer with catalytic core (Figure S11, SI). This observation also clearly defines that not only presence of catalytic core inside the organic network but also suitable orientation of the active sites (in plane & out of plane) governs catalytic efficiency and desired product selectivity accordingly. We have also noticed extreme decline in catalytic performance with our **Fe-Tt-POP** catalyst with the reduction in Fe content during synthesis of metalated POP (Figure S12, SI). Catalytic pocket inside our **Fe-POP** catalysts with proper orientation of neighboring active sites and local binding environment play a crucial role in comparison with the previous reported catalysts which is presented in the respective Table S5, SI. After verifying the catalytic oxidation of TA with both the **Fe-POP** catalysts, the best performing **Fe-Tt-POP** catalyst was employed for the decontamination of different sulfur mustards like 2-Chloroethyl ethyl sulfide (CEES), 2-Chloroethyl methyl sulfide (CEMS), dimethyl sulfide (DMS), ethyl methyl sulfide (EMS) (Figure 4f) under similar reaction conditions. Our as-synthesized **Fe-Tt-POP** catalyst exhibited excellent partial oxidation of sulfur mustard to their corresponding sulfoxide derivatives with almost 95% conversion and more than 78% selectivity for every individual sulfur mustard (Figure 4f). The conversion of CEES was appeared to be poorer when **Fe-Rb-POP** catalyst was used instead of **Fe-Tt-POP** (Figure S13, SI), signifying the decisive role of organic polymer back-bone unit arrangement.

## Mechanistic insights by *in-situ* DRIFTS and Density Functional Theory (DFT) Calculations

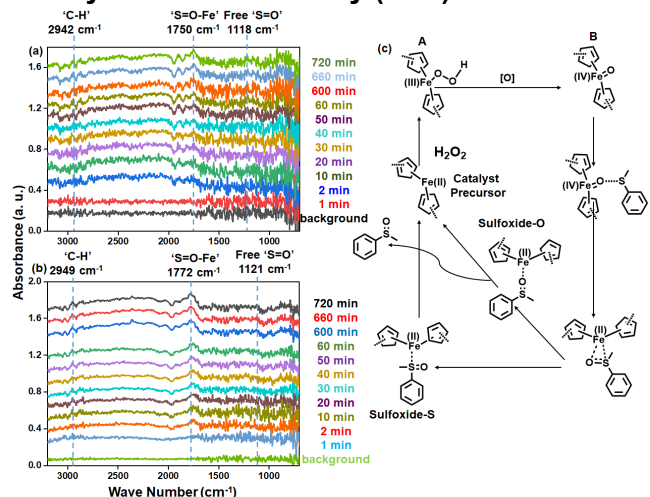

**Figure 5:** *In-situ* DRIFTS measurements of the catalytic oxidation of TA at 100 °C over **Fe-Rb-POP** (a) & **Fe-Tt-POP** (b), respectively. (c) Plausible mechanistic pathway for thioanisole oxidation.

*In situ* DRIFTS was employed to investigate the intermediates formed during the thioanisole oxidation process (Figure 5). It is noted that after 1min reaction time in case of both **Fe-Rb-POP** and **Fe-Tt-POP** sample, the peak at ~1750-1772 cm$^{-1}$, assigned to the most important key intermediate metal coordinated sulfoxide 'S=O-Fe$^{+2}$', whereas the free 'S=O' showed symmetric stretching frequency at ~1120 cm$^{-1}$.[47,48] In the meantime, a 'C-H' symmetric stretching frequency of thioanisole was assigned at ~2945 cm$^{-1}$ for both the catalyst.[49] With the increase of reaction time, the intensity of peak at ~1750-1772 cm$^{-1}$ increased gradually, indicating the formation of key 'S=O-Fe$^{+2}$' intermediate. It is interesting that compare to the free 'S=O' at ~ 1120 cm$^{-1}$, the iron coordinated sulfoxide (Fe-O=S) showed higher stretching frequency. That is ascribed to the transfer of electron from 'O' to Fe$^{+2}$ (d$^6$) center which generates a positive charge density (δ+) on oxygen, making the system highly unstable.[50] Hence, compared to the free 'S=O' of phenyl methyl sulfoxide, a much stronger 'S=O' bond coordinated with Fe with higher stretching frequency was formed to compensate this instability in electron density over sulfur, which is slightly move towards oxygen. More importantly, we observe that the introduction of Tt and Rb leads to different effect on the formation of 'S=O-Fe$^{+2}$' intermediate. As discussed above, the strong 'Fe-O' overlap in the non-planar **Fe-Rb-POP** leads to more shifting of electron density from sulfur to oxygen in the 'S=O-Fe$^{+2}$' intermediate resulting in higher stretching frequency (~1772 cm$^{-1}$), whilst the poor 'Fe-O' overlap in the planar **Fe-Tt-POP** lead to a lower shift of that (~1750 cm$^{-1}$). In order to further confirm the effect of 'Fe-O' overlap on the formation of 'S=O-Fe$^{+2}$' intermediate and affecting the reaction performance, we further carried out the DFT calculations.

In order to shed light towards the catalytic activity of the planar catalytic system (**Fe-Tt-POP**) in comparison to the non-planar one (**Fe-Rb-POP**), we have performed the DFT computational study. As a first step, we have fully relaxed the different modelled systems. Next, we have calculated the free energy for the different reaction intermediates and found that in case of **Fe-Tt-POP**, the value of ΔG is more negative in comparison to the **Fe-Rb-POP**. The more negative value of ΔG depicts more thermodynamic feasibility of **Fe-Tt-POP** (as shown in Figure 6). From Figure 6, we can see that, the value of ΔG for sulfone formation in case of **Fe-Tt-POP** is very negative in comparison to the **Fe-Rb-POP**. Hence, it can be clearly encounter that the large percentage of selectivity has been achieved by the **Fe-Tt-POP** system than the **Fe-Rb-POP** counterpart. Also from the free energy calculation, we have found that the sulfoxide-O pathway is more preferred than the sulfoxide-S pathway in both the **Fe-POPs** case (Table S4, SI).

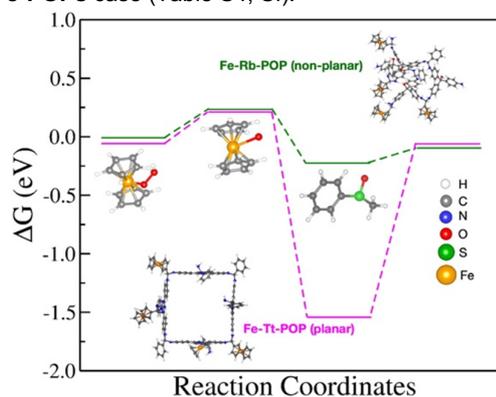

**Figure 6:** Energy profile for sulfoxidation of thioanisole to sulfoxide and sulfone in **Fe-Tt-POP** and **Fe-Rb-POP**.

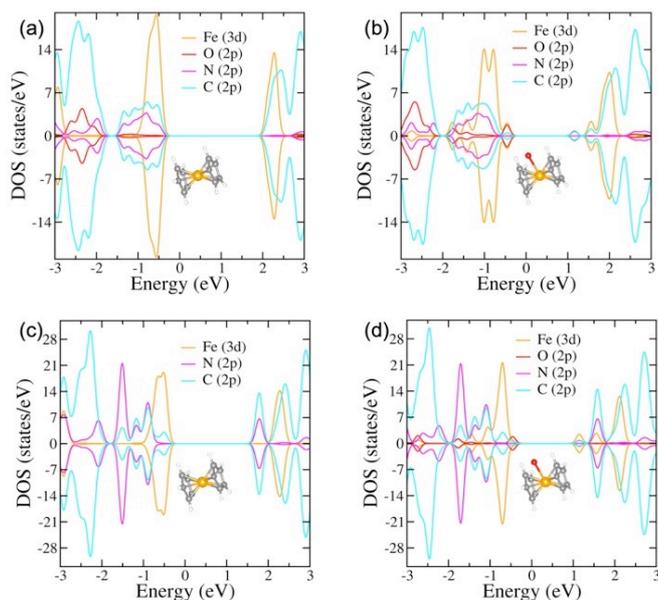

**Figure 7:** Atom-projected partial density of states (PDOS) of (a) **Fe-Rb-POP** without Fe=O, (b) **Fe-Rb-POP** with Fe=O, (c) **Fe-Tt-POP** without Fe=O, and (d) **Fe-Tt-POP** with Fe=O.

Further, to gain better insights, the atom-projected Partial Density of States (PDOS) of two **Fe-POP** systems without 'Fe=O' (Figure 7a & 7c) and with 'Fe=O' (Figure 7b & 7d) are plotted. The PDOS shows, the Fe 3d orbital and C 2p orbital contributes close to the fermi energy level in both the POP systems. Additionally, the PDOS of 'Fe=O' for **Fe-Rb-POP** shows that there is strong hybridization of Fe 3d and O 2p orbital close to the fermi energy level. As a result of intensive hybridization, the 'Fe=O' bond becomes much stronger in the case of the **Fe-Rb-POP**, making it difficult to transfer oxygen from the 'Fe-center' to 'S-center' of thioanisole, resulting in decreased catalytic oxidation. While, in case of **Fe-Tt-POP** there is no such strong interaction between Fe and O making the 'Fe=O' bond more vulnerable. Hence, the 'Fe=O' bond gets dissociated easily in case of **Fe-Tt-POP**, leads to a facile transfer of oxygen atom from 'Fe-center' to 'S-center' of thioanisole showing better catalytic performance. Therefore, from PDOS we can rationalize the reason behind the higher catalytic performance of **Fe-Tt-POP** based system in comparison to the **Fe-Rb-POP** counterpart, which also supports the findings from the energy profile diagram of two **Fe-POP** systems.

## Reusability Test

As the robustness and reusability are two prime factors to evaluate utility of heterogeneous catalysts in industrial scale, we have performed reusability test for our **Fe-Tt-POP** and **Fe-Rb-POP** catalysts up to 8th catalytic run under optimized reaction condition to unveil the heterogeneity. After each catalytic run, the catalysts are simply filtered off and washed by methanol several times and dried in oven at 80º C, then used for the next cycle. The lack of significant drop in catalytic conversion indicates that the heterogeneity remains unaltered in case of both **Fe-Tt-POP** and **Fe-Rb-POP** catalysts during harsh reaction conditions. This demonstrates that the reactivation process is not necessary to activate the catalyst for further catalytic run. In case of our **Fe-Tt-POP** and **Fe-Rb-POP** catalysts, we have observed negligible decline of TA conversion from 94% to 84% and 43% to 37%, respectively even after 8th catalytic cycle (Figure S14, SI). Furthermore, we have also performed PXRD, FT-IR spectroscopy & TEM analysis of the reused **Fe-Tt-POP** and **Fe-Rb-POP** sample to unravel the change in crystallinity and morphology after 8th consecutive catalytic run compared to fresh catalysts. The PXRD of **Fe-Tt-POP** and **Fe-Rb-POP** demonstrate the amorphous nature of porous organic polymer remain unaltered (Figure S15, SI). However, some sharp peaks of lower intensity after 20˚ appeared in both cases, probably due to the formation of iron oxide species at higher temperature, almost unrecognizable with bare eyes.[51] FT-IR spectra of the reused catalysts (Figure S16, SI) clearly illustrate the presence of the secondary amine (N-H) and newly formed methylene (-CH-) linkage on our as-synthesized **Fe-POPs** even after 8th catalytic run. This indicates structural integrity of the highly robust catalysts remain unaltered. TEM analysis of both catalysts provided in Figure S17 in SI, clearly demonstrate encapsulated structure of ferrocenealdehyde based porous organic polymer. The darker and gloomy area defined the presence of metal sites and carbonaceous network of M-POP after 8th catalytic run. For both cases, we have observed random organization & overlapping of different geometric patterns, probably due to the exposure of reactive environment up to 8th catalytic run. Although the morphology of **Fe-Tt-POP** and **Fe-Rb-POP** materials are compromised to some extent, but marginal drop-in catalytic activity even after 8th catalytic cycle towards the oxidative decontamination of sulfur mustards indicates the surface reconstruction and reshaping of the M-POP during reaction.

## Conclusion

Herein, we have successfully designed two **Fe-POPs** having planar & non-planar geometry through ternary copolymerization technique utilizing ferrocenealdehyde (**Fe-CHO**) as metal precursor & two different monomeric units 4,4',4''-(1,3,5-Triazine-2,4,6-triyl) trianiline (**Tt**), Pararosaniline Base (**Rb**) accompanied by benzaldehyde (**Bn-CHO**). The diverse geometry with disorder in local atomic environment was well established by slight change in Debye-Waller factors ($\sigma^2$ value) for **Fe-Rb-POP** compared to **Fe-Tt-POP** from EXAFS analysis. As a consequence of framework structural asymmetry in case of **Fe-Rb-POP**, superior catalytic activity was noticed for **Fe-Tt-POP** counterpart for the oxidative decontamination of sulfur mustards with 99% conversion within 13h. Furthermore, we have conducted some extremely important analysis such as DRIFTS and DFT computational study for the two catalysts to establish the structure and activity relationship. We can finally conclude from *in situ* DRIFTS study that, the very strong overlap between Fe and O in the non-planar **Fe-Rb-POP** able to shift the electron density from sulfur to oxygen in the 'S=O-Fe$^{+2}$' intermediate. We have observed a higher Fe-O stretching frequency (~1772 cm$^{-1}$) shift in comparison with the planar **Fe-Tt-POP** system (~1750 cm$^{-1}$). This outcome indicates the feasible adsorption-desorption of reactant and product during the oxidative decontamination process for **Fe-Tt-POP** system. DFT computational study of both catalysts also revealed that Gibbs free energy of sulfone formation was more negative in case of planar **Fe-Tt-POP** system that facilitates thermodynamically stable product generation compared to non-planar **Fe-Rb-POP** core. In addition, PDOS clearly depicts strong hybridization of Fe 3d and O 2p orbital for **Fe-Rb-POP** system made the oxygen transfer process more difficult from 'Fe-center' to 'S-center' to obtain targeted sulfone product, finely coincide with the above-mentioned results. Hence, we finally tuned the M-POP by introducing planar and non-planar core and also overcome the stability issues of MOFs in heterogeneous catalysis by combining covalent bond and single metal sites for the oxidation of mustard gas simulants which are well-known toxic chemical warfare agent (CWA).

## Acknowledgements

R.P., C.S. and M.J. acknowledge Department of Science & Technology (DST)-INSPIRE (GAP-0799), University Grant Commission (New Delhi) and CSIR, India [Grant No. 09/086(1344)/2018-EMR-I] for their respective senior research fellowships. J.M. acknowledges the Council of Scientific & Industrial Research (CSIR), India, for the CSIR-YSA Research Grant (reference no. HRDG/YSA-19/02/21(0045)/2019) & Focused Basic Research (FBR) Grant under the CLP theme (reference no. 34/1/TD-CLP/NCP-FBR 2020-RPPBDD-TMD-SeMI) for financial support at CSIR-IICT, Hyderabad. S.B. acknowledges the financial support from SERB under core research grant (Grant No. CRG/2019/000647). UK Catalysis Hub is kindly thanked for resources and support provided via our membership of the UK Catalysis Hub Consortium and funded by EPSRC grant: EP/R026939/1, EP/R026815/1, EP/R026645/1, EP/R027129/1 or EP/M013219/1(biocatalysis).

**Keywords:** Metalated Porous-Organic-Polymer; Ternary copolymerization; Catalytic pocket; Mustard-Gas Stimulant; System planarity.

**Entry for the Table of Contents**

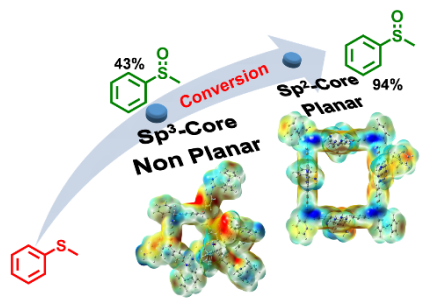

Core planarity driven detoxification of mustard gas simulant, whereas planar **Fe-Tt-POP** exhibited superior catalytic performance compared to non-planar **Fe-Rb-POP**. The observed difference in catalytic performance could be attributed to the proper alignment & orientation of catalytic pocket inside the porous framework.

# Metalated Porous-Organic-Polymer Renders Mustard-Gas Simulant Harmless: Core Planarity Matters

Ratul Paul,[a, b], ‡ Chitra Sarkar,[a, b], ‡ Manjari Jain,[c], ‡ Shaojun Xu,[d, e] Kashmiri Borah,[f] Duy Quang Dao,[g] Chih-Wen Pao,[h] Saswata Bhattacharya*,[c] & John Mondal*,[a, b]

[a]     R. Paul, C. Sarkar and Dr. J. Mondal
Catalysis & Fine Chemicals Division, CSIR-Indian Institute of Chemical Technology, Uppal Road, Hyderabad 500 007, India. johnmondal@iict.res.in; Email: johnmondal@iict.res.in, johncuchem@gmail.com (J.M.)
[b]     R. Paul, C. Sarkar and Dr. J. Mondal
        Academy of Scientific and Innovative Research (AcSIR), Ghaziabad- 201002, India.
[c]     M. Jain and Dr. S. Bhattacharya
Department of Physics, Indian Institute of Technology Delhi Hauz Khas, New Delhi 110 016, India.
Email: saswata@physics.iitd.ac.in  (S.B.)
[d]     Dr. S. Xu
Cardiff Catalysis Institute, School of Chemistry, Cardiff University, Cardiff CF10 3AT, U.K.
[e]     Dr. S. Xu
        UK Catalysis Hub, Research Complex at Harwell, Rutherford Appleton Laboratory, Harwell, OX11 0FA, United Kingdom.
[f]     K. Borah
Polymers & Functional Materials Division, CSIR-Indian Institute of Chemical Technology, Uppal Road, Hyderabad 500 007, India.
[g]     Dr. D. Q. Dao
Institute of research and development, Duy Tan University, Da Nang, 550000, Viet Nam.
[h]     Dr. C. W. Pao
National Synchrotron Radiation Research Center, 101 Hsin-Ann Road, Hsinchu 30076, Taiwan.
‡       R.P., C.S. and M.J. are equally contributed to this work.

## Table of content

**Characterization Details:**

Powder X-ray Diffraction pattern (PXRD) of different samples were recorded with a Bruker D8 Advance X-ray diffractometer operated at a voltage of 40 kV and a current of 40 mA by using Ni-filtered Cu Kα (l = 0.15406 nm) radiation. High-resolution Transmission Electron Microscopy (HR-TEM) images were recorded with a JEOL JEM 2010 transmission electron microscope with operating voltage 200 kV, equipped with a Field-emission Electron Gun (FEG). Field-emission scanning electron microscopic images of samples were obtained by using a JEOL JEM 6700 Field-emission Scanning Electron Microscope (FE-SEM). Nitrogen sorption isotherms were obtained by using a Quantachrome Autosorb 1C surface area analyzer at 77 K. Prior to the measurements, the samples were degassed at 393 K for approximately 4 h under high vacuum. Surface areas were calculated from the adsorption data by using the Brunauer-Emmett-Teller (BET) method in the relative pressure (P/P0) range 0.01-0.1. The total pore volumes and pore size distribution curves were obtained from the adsorption branches by using Nonlocal Density Functional Theory (NLDFT) method. Fourier Transform Infrared Spectroscopy (FTIR) spectra of the samples were recorded by using a Nicolet MAGNA-FT IR 750 Spectrometer Series II. Solid-state Cross-Polarization Magic Angle Spinning Carbon-13 Nuclear Magnetic Resonance ($^{13}$C CP MAS NMR) studies were performed by using a Bruker Advance III HD 400 MHz NMR spectrometer. High-angle Annular Dark-field Scanning Transmission Electron Microscopy (HAADF-STEM) and energy-dispersive X-ray mapping images were obtained with a TECNAI G2 F20 equipped with an EDX detector. X-ray Photoelectron Spectroscopy (XPS) was performed with an Omicron nanotech operated at 15 kV and 20 mA with a monochromatic Al Kα X-ray source. The Fe K-edge X-ray Absorption Spectra (XAS) were measured using beamline 44A at Taiwan Photon Source (TPS) of the National Synchrotron Radiation Research Centre (NSRRC, Hsinchu, Taiwan). All spectra were performed in transmission mode at room temperature. The incoming and outgoing photon fluxes were measured by ionization chambers filled with appropriate mixtures of $N_2$ and Kr gases. For these measurements, the nanohybrids were uniformly diluted inside an inert boron nitride matrix and pressed to a form of pellet to have absorption edge jump between 0.5 and 1.

All calculations were performed using Gaussian 16 Rev.A.03.[1] Optimized geometries and frequencies were calculated at the Becke-3 Parameter-Lee-Yang-Parr hybrid functionals B3LYP[2] combined with the Def2-SV(P) basis set.[3,4] This basis set was successfully employed by Jorge Gutierrez-Flores *et.al* (2020) to evaluate the stabilization of Tominaga's

M12L24 nanoballs with Pd$^{2+}$ and Ni$^{2+}$ as metal centers.[5] Frontier molecular orbital distributions including highest occupied molecular orbital (HOMO) and lowest unoccupied molecular orbital (LUMO) and electrostatic potential (ESP) maps are also investigated using Gauss View 6 to predict the electronic properties of the Porous Organic Polymer (POP) units. Density Functional Theory (DFT) calculations were performed using the Vienna *ab initio* Simulation Package (VASP)[6] with the Projected Augmented Wave (PAW) potential.[7] The Perdew–Burke-Ernzerhof (PBE)[8] exchange-correlation functional within the Generalized Gradient Approximation (GGA) was used in all the calculations. The cut-off energy of 500 eV was chosen for the plane-wave basis set. The tolerance energy for the convergence was set to 0.001 meV to achieve self-consistency in the total energy. All the electronic configurations were fully relaxed until the ionic forces were smaller than 0.01 eV/Å using conjugate gradient minimization. Their corresponding binding energies direct towards the type of adsorption in the system.

**Experimental Section:**

**Materials:**

All the required chemicals ferrocenecarboxaldehyde, pararosaniline base, 4,4',4"-(1,3,5-Triazine-2,4,6-triyl) trianiline, benzaldehyde, 2-Chloroethyl ethyl sulfide (CEES), 2-Chloroethyl methyl sulfide (CEMS), dimethyl sulfide (DMS), ethyl methyl sulfide (EMS) and hydrogen peroxide were purchased from Sigma-Aldrich and used as received unless noted otherwise. Dimethyl sulfoxide (DMSO), methanol (MeOH), acetonitrile solvents were all dried before using in reaction.

**Synthesis of Pararosaniline based Porous-Organic Polymer (Fe-Rb-POP):**

In a typical synthetic procedure ferrocenecarboxaldehyde (2 mmol, 0.41 gm), Pararosaniline Base (4 mmol, 1.22 gm) and Benzaldehyde (2 mmol, 0.102 ml) were mixed with DMSO (12 ml) in a 100 mL two-neck round bottom flask equipped with a stirrer and a condenser. Then, the reaction mixture was stirred under N$_2$ atmosphere for 1 hour to get a homogeneous mixture. Further, the prepared solution was taken in a seal tube and heated at 140º C for 40 h. Finally, the mixture was cooled down to room-temperature and the product was collected and washed with MeOH for several times and dried in a vacuum oven at 100º C overnight to get the blackish powder.

**Synthesis of 4,4',4''-(1,3,5-Triazine-2,4,6-triyl) trianiline based Porous-Organic Polymer (Fe-Tt-POP):**

Similar synthetic protocol was followed as mentioned above, but instead of Pararosaniline Base, here we use 4,4',4''-(1,3,5-Triazine-2,4,6-triyl) trianiline (2 mmol, 0.708 g).

**Liquid-phase Selective Oxidation of Sulfides to Sulfoxides:**

Liquid phase oxidation of sulfides to sulfoxides was performed using Thioanisole (TA) as a model sulfide compound. At first, a well-dispersed mixture of 0.1 ml of TA (0.5 mmol), Fe-POP catalyst (50 mg), and $H_2O_2$ (0.25 ml) was heated at 100° C for desired time. After the completion of the reaction, the catalyst was separated from the reaction mixture via filtration. Further, the collected reaction mixture was analyzed by a gas chromatograph (Shimadzu 2010) equipped with a flame ionization detector using an INNOWax capillary column (diameter: 0.25 mm, length: 30 m). The products were also identified by GC-MS (Shimadzu, GCMS-QP2010S). After the confirmation of sulfide to sulfoxide conversion, catalytic oxidation of other sulfur mustard (HD) compounds was carried out under similar reaction conditions to convert HD's into less toxic sulfoxide derivatives.

**_In situ_ Diffuse Reflectance Infrared Fourier Transform Spectroscopy (DRIFTS):**

The _in-situ_ DRIFTS experiment was carried out using an Agilent Carey 680 FTIR Spectrometer equipped with a Harrick DRIFTS cell. The spectra were recorded at 4 cm$^{-1}$ resolution and each spectrum was averaged 64 times. 50 mg of sample was first mixed with 0.12 ml (0.1 mmol) of thioanisole and 0.25 ml $H_2O_2$ using an incipient wetness impregnation method to form a solid-liquid mixture. Then the incipiently wetted sample was placed into the sample holder in the Harrick DRIFTS cell. Each sample was pre-treated at 30° C using 99.999% argon (Ar, BOC gas Ltd.) at a gas flow rate of 50 ml min$^{-1}$ for 1h in order to remove the residual air and impurities in the cell. Then the purged sample was taken as the background spectrum. After that the temperature of the sample was rapidly increased to 100° C at a 50° C/min rate under the same Ar gas condition, then maintained at 100° C for 13h to determine the surface intermedia revolution. The sample **Fe-Rb-POP** and **Fe-Tt-POP** was tested using the operando DRIFTS.

**Table S1: Elemental C, H, N analysis**

| Catalyst | Nitrogen (wt%) | Carbon (wt%) | Hydrogen (wt%) |
|---|---|---|---|
| **Fe-Rb-POP** | 6.254 | 72.646 | 14.249 |
| **Fe-Tt-POP** | 10.947 | 72.232 | 12.471 |

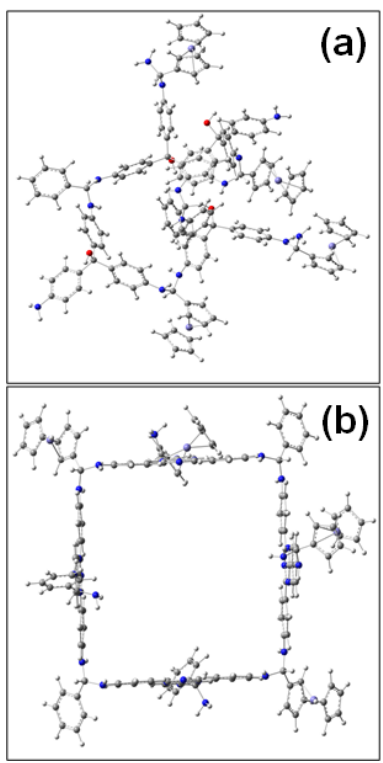

**Figure S1:** Optimized geometry of the **Fe-Rb-POP** (a) and **Fe-Tt-POP** (b) systems, respectively.

**Table S2:** Cartesian coordinates and thermochemical properties of two POP structures calculated in the gas phase at the B3LYP/Def2-SV(P) level of theory.

| Fe-Tt-POP | | | |
|---|---|---|---|
| 0 1 | | | |
| C | -2.76974500 | 8.27339000 | 4.21654900 |
| N | -3.66221800 | 7.93935600 | 3.26058000 |
| C | -4.28857300 | 6.75555100 | 3.42692200 |
| C | -3.17005100 | 6.34406100 | 5.37875300 |
| N | -2.49565000 | 7.50977000 | 5.29574100 |
| N | -4.07023600 | 5.92225600 | 4.46598900 |
| C | -5.28281700 | 6.34892700 | 2.41052600 |
| C | -5.95848800 | 5.10890200 | 2.50255600 |
| C | -5.59129900 | 7.18759300 | 1.31770400 |
| C | -6.90150200 | 4.73067300 | 1.55032700 |
| H | -5.72088900 | 4.44384500 | 3.34685400 |
| C | -6.53609300 | 6.81852300 | 0.35722000 |
| H | -5.06443700 | 8.15066300 | 1.23476600 |
| C | -7.22521400 | 5.58045700 | 0.45751100 |
| H | -7.41584300 | 3.75839500 | 1.64272200 |
| H | -6.73969800 | 7.50990800 | -0.47544600 |
| C | -2.91264200 | 5.47637900 | 6.54888400 |
| C | -3.56154200 | 4.22689200 | 6.69005500 |
| C | -2.01401100 | 5.87037600 | 7.56352700 |
| C | -3.32678700 | 3.41519100 | 7.79707800 |
| H | -4.25917500 | 3.90633300 | 5.90131800 |
| C | -1.77147600 | 5.06407000 | 8.67839500 |
| H | -1.49984600 | 6.83801200 | 7.45755800 |
| C | -2.43471800 | 3.81749200 | 8.82782900 |
| H | -3.84431500 | 2.44425200 | 7.88563600 |
| H | -1.06166600 | 5.41967900 | 9.44169500 |
| N | -2.27892600 | 3.00434200 | 9.94261800 |
| H | -2.65152800 | 2.05604600 | 9.84432200 |
| C | -1.21603700 | 3.15237300 | 10.92160800 |
| H | -1.14749000 | 4.23826100 | 11.15301400 |
| N | 0.12230500 | 2.79985500 | 10.44228300 |
| C | 0.47045300 | 1.57054100 | 9.88717600 |
| H | 0.63811500 | 3.58760700 | 10.04346800 |
| C | 1.64009800 | 1.48504900 | 9.08544200 |
| C | -0.26875300 | 0.37914300 | 10.11026100 |
| C | 2.05804700 | 0.27019400 | 8.54654900 |
| H | 2.22530500 | 2.40081100 | 8.89031300 |
| C | 0.15663200 | -0.83301900 | 9.55978800 |
| H | -1.16904000 | 0.39228100 | 10.74245600 |
| C | 1.32537300 | -0.91831900 | 8.77281200 |
| H | 2.96772300 | 0.21611300 | 7.92920100 |
| H | -0.41877300 | -1.75399900 | 9.74034200 |
| C | 1.77422100 | -2.21084000 | 8.20820600 |
| N | 1.02244100 | -3.30332000 | 8.45529600 |
| N | 2.91317700 | -2.22125200 | 7.48490500 |
| C | 1.47551900 | -4.46149400 | 7.92869900 |
| C | 3.29258000 | -3.42399600 | 7.00258500 |
| N | 2.60671600 | -4.56892300 | 7.19935400 |
| C | 4.54240900 | -3.49036500 | 6.21442600 |
| C | 5.30062000 | -2.32593800 | 5.94902400 |
| C | 5.02032600 | -4.71794700 | 5.70612100 |
| C | 6.48001300 | -2.38970400 | 5.21068300 |
| H | 4.93530400 | -1.36232300 | 6.33577400 |
| C | 6.20502700 | -4.79490600 | 4.96939700 |
| H | 4.43616100 | -5.62882300 | 5.90854500 |
| C | 6.96750100 | -3.62587700 | 4.70635900 |
| H | 7.05163800 | -1.46626500 | 5.01149100 |
| H | 6.55221500 | -5.77713900 | 4.61604500 |
| N | 8.17201600 | -3.63965500 | 4.01060900 |
| H | 8.49471800 | -2.71025400 | 3.73100700 |
| C | 8.62258700 | -4.71314700 | 3.11821600 |
| H | 9.56953600 | -4.32895800 | 2.68779400 |
| N | 7.74175200 | -5.00448400 | 2.00003600 |
| H | 6.94793000 | -5.61241600 | 2.21772600 |
| C | -1.58637000 | 2.46082000 | 12.24134300 |
| C | -0.57761900 | 2.03031500 | 13.12604700 |
| C | -2.93471600 | 2.31934900 | 12.62681000 |
| C | -0.91005600 | 1.46033400 | 14.36410400 |
| H | 0.47682600 | 2.13968300 | 12.82710800 |
| C | -3.26697600 | 1.74746800 | 13.86625800 |
| H | -3.73121400 | 2.67100200 | 11.95169500 |
| C | -2.25624600 | 1.31444300 | 14.73864600 |
| H | -0.10779000 | 1.12223000 | 15.04088200 |
| H | -4.32719300 | 1.64369000 | 14.15140500 |
| H | -2.51676400 | 0.86353000 | 15.71057000 |
| C | -2.04782600 | 9.55515700 | 4.07279900 |
| C | -1.09661400 | 9.96835300 | 5.03132300 |
| C | -2.28067900 | 10.40592300 | 2.96570700 |
| C | -0.40217500 | 11.17318900 | 4.90133200 |
| H | -0.90498200 | 9.30962200 | 5.89237300 |
| C | -1.59789400 | 11.61120600 | 2.82619300 |
| H | -3.01751900 | 10.09193800 | 2.21054100 |
| C | -0.64165800 | 12.03083800 | 3.79261700 |
| H | 0.33716000 | 11.44712300 | 5.67012000 |
| H | -1.79904700 | 12.26044300 | 1.95654900 |
| C | 0.68479700 | -5.68639600 | 8.16651900 |
| C | 1.09193600 | -6.93205300 | 7.63973600 |
| C | -0.50798100 | -5.65109200 | 8.92883300 |
| C | 0.35530200 | -8.09764600 | 7.86117300 |
| H | 2.01450600 | -6.96695100 | 7.03989700 |
| C | -1.25168400 | -6.80491200 | 9.15783900 |
| H | -0.83686800 | -4.68461600 | 9.34073100 |
| C | -0.84149100 | -8.06311100 | 8.63139100 |
| H | 0.70834900 | -9.04037000 | 7.41465700 |
| H | -2.17809700 | -6.75240400 | 9.75614800 |
| N | -0.00763700 | 13.25124700 | 3.63687500 |
| H | -0.07851800 | 13.67853900 | 2.70947000 |
| N | -1.60196900 | -9.18579600 | 8.87337100 |
| H | -2.46383800 | -9.05802700 | 9.40715500 |
| C | 1.14010100 | 13.68449100 | 4.41604000 |
| H | 0.86963000 | 13.49476200 | 5.48657200 |
| C | -1.25182200 | -10.55641300 | 8.47619700 |
| H | -0.95567200 | -10.52623100 | 7.40389300 |
| N | 2.34816800 | 12.97584900 | 3.98392300 |
| H | 3.14198900 | 13.20834900 | 4.60058200 |
| H | 2.19385500 | 11.95829700 | 4.04495900 |
| N | -0.13823600 | -11.17109400 | 9.19164800 |
| H | 0.69913100 | -10.57229200 | 9.12955400 |
| H | -0.36632000 | -11.26608200 | 10.19275700 |
| C | 7.70219000 | -0.08810200 | -4.05112000 |
| N | 7.89362900 | -0.78913500 | -2.91265900 |
| C | 7.03516400 | -1.80673600 | -2.69540600 |
| C | 5.92272700 | -1.39584900 | -4.65011000 |
| N | 6.73227600 | -0.36005100 | -4.95114400 |
| N | 6.03097600 | -2.14406500 | -3.53193900 |
| C | 7.21009700 | -2.61002700 | -1.46557300 |
| C | 6.32929500 | -3.67268800 | -1.15380200 |
| C | 8.26453900 | -2.34321000 | -0.56593400 |
| C | 6.50087800 | -4.43522700 | -0.00118300 |
| H | 5.49919100 | -3.88519500 | -1.84473500 |
| C | 8.44662100 | -3.10243600 | 0.59259100 |
| H | 8.94925400 | -1.51279500 | -0.79739700 |
| C | 7.57205400 | -4.17921900 | 0.89763400 |
| H | 5.80128900 | -5.25962300 | 0.22114000 |
| H | 9.28540200 | -2.85211100 | 1.26094900 |
| C | 4.85130400 | -1.73969700 | -5.61105200 |
| C | 3.95146700 | -2.80102300 | -5.35481300 |
| C | 4.69646400 | -1.02104300 | -6.81584700 |
| C | 2.95012400 | -3.13105500 | -6.26493300 |
| H | 4.05856800 | -3.36366400 | -4.41484600 |
| C | 3.69525500 | -1.34338800 | -7.73546500 |
| H | 5.38842800 | -0.18928800 | -7.01903500 |
| C | 2.80288200 | -2.41913400 | -7.48599200 |
| H | 2.25949400 | -3.96324300 | -6.04367400 |
| H | 3.61933900 | -0.75119400 | -8.66096800 |
| N | 1.83017400 | -2.82336000 | -8.39165500 |
| H | 1.13346000 | -3.47375900 | -8.01827000 |
| C | 1.38745000 | -2.02841000 | -9.52358800 |
| H | 2.30525200 | -1.62928700 | -10.00893500 |
| N | 0.61307600 | -0.83379700 | -9.17735400 |
| C | -0.54895100 | -0.83514300 | -8.40951800 |
| H | 1.18435200 | 0.00486200 | -9.05053200 |
| C | -0.96497700 | 0.37312400 | -7.78866900 |
| C | -1.36772500 | -1.98280500 | -8.24102000 |
| C | -2.14533900 | 0.43606400 | -7.05122900 |
| H | -0.33853300 | 1.27582400 | -7.89738100 |
| C | -2.54641500 | -1.90967600 | -7.49386600 |
| H | -1.09771600 | -2.93507900 | -8.72120100 |
| C | -2.96480300 | -0.70523900 | -6.88802800 |
| H | -2.45868600 | 1.37911100 | -6.57787700 |
| H | -3.17910800 | -2.80218400 | -7.37144900 |
| C | -4.22506800 | -0.63669200 | -6.11528200 |
| N | -4.95097200 | -1.76686000 | -5.98967400 |
| N | -4.57903500 | 0.55417600 | -5.58886200 |
| C | -6.10003100 | -1.65442600 | -5.28947000 |
| C | -5.74506900 | 0.57341500 | -4.90845900 |
| N | -6.53799700 | -0.50437100 | -4.73436400 |
| C | -6.18411700 | 1.85963800 | -4.32494700 |
| C | -5.39779300 | 3.02798500 | -4.45885100 |
| C | -7.40425300 | 1.96351800 | -3.62203800 |
| C | -5.81657600 | 4.24133100 | -3.91803500 |
| H | -4.44147700 | 2.95790400 | -4.99933400 |
| C | -7.83672000 | 3.17626700 | -3.07940100 |
| H | -8.02133400 | 1.05851300 | -3.51206500 |
| C | -7.05044300 | 4.35052900 | -3.22100500 |
| H | -5.18372000 | 5.13889700 | -4.03222800 |
| H | -8.80668800 | 3.21437900 | -2.56193300 |
| N | -7.44260100 | 5.59606900 | -2.74160400 |
| H | -6.69281500 | 6.29149200 | -2.74214000 |
| C | -8.46398600 | 5.82742800 | -1.71571600 |
| H | -8.43169800 | 6.92216300 | -1.54055800 |
| N | -8.21455900 | 5.19050800 | -0.43296200 |
| H | -8.49970600 | 4.20979400 | -0.36997700 |
| C | 0.70037800 | -2.90871600 | -10.57713500 |
| C | -0.22717300 | -2.34778400 | -11.47751000 |
| C | 1.04477200 | -4.26850700 | -10.71478800 |
| C | -0.80628400 | -3.13223900 | -12.48618600 |
| H | -0.49799500 | -1.28536400 | -11.37202200 |
| C | 0.46368000 | -5.05378600 | -11.72441200 |
| H | 1.78403100 | -4.71224900 | -10.02893500 |
| C | -0.46501400 | -4.48921200 | -12.61263900 |
| H | -1.53584700 | -2.67858600 | -13.17750600 |
| H | 0.74407800 | -6.11629200 | -11.81716900 |
| H | -0.92296200 | -5.10584300 | -13.40376700 |
| C | 8.60730300 | 1.04504000 | -4.32757600 |
| C | 8.47470000 | 1.81774000 | -5.50287500 |
| C | 9.63939300 | 1.39592700 | -3.42382100 |
| C | 9.32455400 | 2.89167600 | -5.77545600 |
| H | 7.68025900 | 1.54819400 | -6.21586900 |
| C | 10.48762600 | 2.46907200 | -3.67892400 |
| H | 9.75410700 | 0.80165200 | -2.50428500 |
| C | 10.35212100 | 3.25509000 | -4.86016400 |
| H | 9.19786800 | 3.44848600 | -6.71633200 |
| H | 11.28026200 | 2.72711000 | -2.95509500 |
| C | -6.93051300 | -2.86558200 | -5.12475000 |
| C | -8.15103100 | -2.82467200 | -4.41469700 |
| C | -6.53020600 | -4.10564700 | -5.67811000 |
| C | -8.94778700 | -3.96170200 | -4.26220200 |
| H | -8.46744600 | -1.86480700 | -3.97797100 |
| C | -7.31546300 | -5.24584800 | -5.53299300 |
| H | -5.58043900 | -4.14838500 | -6.23304900 |
| C | -8.54634300 | -5.20773600 | -4.81926700 |
| H | -9.88458200 | -3.88321800 | -3.68837700 |
| H | -6.98428200 | -6.20235300 | -5.97421100 |
| N | 11.21927400 | 4.30541800 | -5.06726400 |
| H | 11.84776000 | 4.54245600 | -4.29368400 |
| N | -9.28770000 | -6.36441800 | -4.67619500 |
| H | -8.98789500 | -7.16665000 | -5.24022800 |
| C | 11.09104700 | 5.39466000 | -6.02447300 |
| H | 10.10958100 | 5.26666700 | -6.54108200 |
| C | -10.70910900 | -6.40622900 | -4.31344600 |
| H | -10.81785200 | -5.84499800 | -3.35624800 |
| N | 11.18500500 | 6.64428300 | -5.25935100 |
| H | 11.40501000 | 7.43377400 | -5.88235300 |
| H | 10.29210700 | 6.84650200 | -4.78516800 |
| N | -11.64444000 | -5.80345300 | -5.25344000 |
| H | -11.40842500 | -4.81264000 | -5.40818200 |
| H | -11.58304200 | -6.28776800 | -6.16198400 |
| C | -11.08529500 | -7.84687600 | -4.05255200 |
| C | -10.23257100 | -8.86609800 | -3.48701000 |
| C | -12.38097100 | -8.43535500 | -4.28125800 |
| H | -9.18272100 | -8.73161000 | -3.19341700 |
| C | -11.00040200 | -10.07646800 | -3.37307600 |
| H | -13.24226400 | -7.90343800 | -4.70482300 |
| C | -12.32685900 | -9.81028400 | -3.86540400 |
| H | -10.63322700 | -11.03887600 | -2.98976800 |
| H | -13.15040000 | -10.53556800 | -3.92615600 |
| Fe | -10.93323500 | -9.45497500 | -5.31936400 |
| C | -9.32371600 | -9.70593400 | -6.56858300 |
| C | -9.97818200 | -10.96278500 | -6.31578800 |
| C | -10.28678400 | -8.81545100 | -7.16548800 |
| H | -8.26943700 | -9.47679500 | -6.35533000 |
| H | -9.51681900 | -11.85032100 | -5.86069900 |
| C | -11.34456100 | -10.84901400 | -6.75469700 |
| H | -10.09320400 | -7.78625500 | -7.50162200 |
| C | -11.53704900 | -9.52112900 | -7.27584200 |
| H | -12.11092200 | -11.63411500 | -6.69082600 |
| H | -12.47397700 | -9.11753200 | -7.68474700 |
| C | 12.17699200 | 5.41124200 | -7.08400200 |
| C | 12.07490800 | 6.06606400 | -8.36907200 |
| C | 13.53910400 | 4.96347900 | -6.93259400 |
| H | 11.16231100 | 6.51009000 | -8.79362800 |
| C | 13.36352900 | 6.01001500 | -9.00620300 |
| H | 13.93694500 | 4.41237000 | -6.06968100 |
| C | 14.26740300 | 5.32777900 | -8.11748100 |
| H  | 13.60807500  | 6.39751200  | -10.00536900 |
| H  | 15.32333400  | 5.10057100  | -8.32112800 |
| Fe | 12.67816600  | 4.12889200  | -8.59083900 |
| C  | 11.12287100  | 2.96258200  | -9.24756900 |
| C  | 12.00803500  | 3.29004300  | -10.33407300 |
| C  | 11.88360600  | 2.26381400  | -8.24574500 |
| H  | 10.05259200  | 3.20663800  | -9.19241600 |
| H  | 11.73627200  | 3.83542800  | -11.24877800 |
| C  | 13.31735700  | 2.79218300  | -10.00361800 |
| H  | 11.49545100  | 1.89139800  | -7.28761900 |
| C  | 13.23940000  | 2.16021700  | -8.71307300 |
| H  | 14.22197500  | 2.89505800  | -10.61935700 |
| H  | 14.07412300  | 1.69552800  | -8.16991300 |
| C  | 8.94075400   | -5.99519300 | 3.87697700 |
| C  | 8.75284400   | -7.34930100 | 3.40802500 |
| C  | 9.48779900   | -6.06590500 | 5.20915700 |
| H  | 8.37927700   | -7.63866700 | 2.41666700 |
| C  | 9.17375000   | -8.24562600 | 4.45273400 |
| H  | 9.73799800   | -5.20086100 | 5.83742000 |
| C  | 9.62684300   | -7.45201400 | 5.56386200 |
| H  | 9.16696800   | -9.34332000 | 4.40072200 |
| H  | 10.02505300  | -7.83607200 | 6.51328900 |
| H  | 12.36828100  | -4.74141700 | 3.56619600 |
| C  | 12.30609800  | -5.82871700 | 3.41521900 |
| C  | 11.81409700  | -6.49629800 | 2.23863600 |
| C  | 12.69047100  | -6.83517800 | 4.37029700 |
| H  | 11.42729000  | -6.01140300 | 1.33064400 |
| C  | 11.89630300  | -7.91544300 | 2.46608200 |
| H  | 13.09064700  | -6.65108200 | 5.37712000 |
| C  | 12.43927600  | -8.12440900 | 3.78221800 |
| H  | 11.58225400  | -8.70022200 | 1.76379400 |
| H  | 12.61146700  | -9.09771300 | 4.26275000 |
| Fe | 10.71191800  | -7.02670600 | 3.87963500 |
| C  | -9.86117500  | 5.47072500  | -2.20691800 |
| C  | -10.95800000 | 4.97653500  | -1.40617900 |
| C  | -10.32132300 | 5.56176200  | -3.57039800 |
| H  | -10.93873400 | 4.82184200  | -0.31903900 |
| C  | -12.08222500 | 4.75549000  | -2.27745100 |
| H  | -9.71614500  | 5.90158400  | -4.42112700 |
| C  | -11.68732600 | 5.11607800  | -3.61304900 |
| H  | -13.07310000 | 4.39271200  | -1.97045600 |
| H  | -12.32245700 | 5.07458500  | -4.50881200 |
| H  | -9.98054500  | 9.05769500  | -2.87437100 |
| C  | -10.93917000 | 8.67785500  | -2.49275500 |
| C  | -11.23420800 | 8.33641300  | -1.12585500 |
| C  | -12.11655900 | 8.42022000  | -3.28012600 |
| H  | -10.54057500 | 8.40446700  | -0.27530700 |
| C  | -12.59457200 | 7.86904900  | -1.06845400 |
| H  | -12.21174300 | 8.56304800  | -4.36559900 |
| C  | -13.13981900 | 7.92252400  | -2.39909100 |
| H  | -13.11803100 |  7.51686900 | -0.16863300 |
| H  | -14.15310200 |  7.61631500 | -2.69466600 |
| Fe | -11.48864600 |  6.71097100 | -2.34429700 |
| C  |   1.30695200 | 15.19624000 |  4.26224100 |
| C  |   0.48173500 | 16.06796100 |  5.00068700 |
| C  |   2.23730900 | 15.73669900 |  3.35522000 |
| C  |   0.58743200 | 17.45704600 |  4.83947300 |
| H  |  -0.25351800 | 15.64795900 |  5.70869400 |
| C  |   2.34516000 | 17.12893200 |  3.19407400 |
| H  |   2.87681700 | 15.04443800 |  2.78415500 |
| C  |   1.52162800 | 17.99165600 |  3.93429600 |
| H  |  -0.06136400 | 18.12893000 |  5.42582300 |
| H  |   3.07996900 | 17.54235100 |  2.48285400 |
| H  |   1.60747900 | 19.08398400 |  3.80893500 |
| C  |  -2.49275500 | -11.43121400 |  8.60172800 |
| C  |  -3.14081100 | -11.58108700 |  9.84659000 |
| C  |  -3.00808400 | -12.11887500 |  7.48699100 |
| C  |  -4.27683600 | -12.39655600 |  9.97141200 |
| H  |  -2.75523500 | -11.05905800 | 10.74062800 |
| C  |  -4.14331700 | -12.93576400 |  7.60841100 |
| H  |  -2.50964800 | -12.01292800 |  6.50868200 |
| C  |  -4.78181600 | -13.07665400 |  8.85124700 |
| H  |  -4.76931400 | -12.50284100 | 10.95232400 |
| H  |  -4.53259000 | -13.46608400 |  6.72348200 |
| H  |  -5.67376900 | -13.71758400 |  8.94771600 |

| | |
|---|---|
| Zero-point correction =                        | 2.504931 (Hartree/Particle) |
| Thermal correction to Energy =                 | 2.666607 |
| Thermal correction to Enthalpy =               | 2.667552 |
| Thermal correction to Gibbs Free Energy =      | 2.260295 |
| Sum of electronic and zero-point Energies =    | -12593.363526 |
| Sum of electronic and thermal Energies =       | -12593.201849 |
| Sum of electronic and thermal Enthalpies =     | -12593.200905 |
| Sum of electronic and thermal Free Energies =  | -12593.608161 |
| HF= -12595.8684566 | |

**Fe-Rb-POP**

0 1

| | | | |
|---|---|---|---|
| C | -1.27558000 | -2.43690500 |  0.77230200 |
| C | -0.89866700 | -3.87578600 |  0.38508600 |
| C | -1.04653600 | -4.97763700 |  1.24664400 |
| C | -0.44626600 | -4.13479100 | -0.92838800 |
| C | -0.74718700 | -6.28643400 |  0.83690600 |
| H | -1.39319100 | -4.81747600 |  2.27999800 |
| C | -0.15552100 | -5.43242500 | -1.35446900 |
| H | -0.33153400 | -3.29848800 | -1.63780700 |
| C | -0.30317500 | -6.54466700 | -0.48207100 |
| H | -0.81909800 | -7.10823400 |  1.56296800 |
| H |  0.16925200 | -5.60348700 | -2.39555500 |
| C | -2.54668600 | -2.04113600 | -0.00461300 |
| C | -3.69618100 | -2.85637500 | 0.04396300 |
| C | -2.62449500 | -0.86349800 | -0.77011300 |
| C | -4.88326400 | -2.50613900 | -0.61226800 |
| H | -3.66487400 | -3.80370800 | 0.60778500 |
| C | -3.80367900 | -0.50150900 | -1.43421900 |
| H | -1.74282200 | -0.21127600 | -0.85889800 |
| C | -4.96458200 | -1.30781000 | -1.36774800 |
| H | -5.74379200 | -3.18991800 | -0.56150800 |
| H | -3.83002400 | 0.43658300 | -2.01581900 |
| C | -1.43729500 | -2.23893200 | 2.28800100 |
| C | -0.32693500 | -2.47784600 | 3.12930800 |
| C | -2.61284400 | -1.75622100 | 2.89978500 |
| C | -0.37843200 | -2.26278900 | 4.51025600 |
| H | 0.61666900 | -2.83719700 | 2.68659800 |
| C | -2.68173200 | -1.54061300 | 4.28090400 |
| H | -3.49902800 | -1.53798500 | 2.28396500 |
| C | -1.56998200 | -1.79607700 | 5.12194000 |
| H | 0.52734700 | -2.44538500 | 5.10708700 |
| H | -3.62012400 | -1.16440700 | 4.72468400 |
| O | -0.14955900 | -1.59916500 | 0.35774300 |
| H | -0.12667200 | -0.83238500 | 0.97273700 |
| N | -6.12315500 | -0.91880200 | -2.05628600 |
| N | -1.68880500 | -1.60552000 | 6.49328800 |
| C | -7.47264800 | -1.18800800 | -1.52809000 |
| H | -8.14191200 | -0.56901900 | -2.15715500 |
| N | -7.68024900 | -0.76588300 | -0.14129200 |
| C | -0.60543100 | -1.66907100 | 7.46009800 |
| H | -0.01243900 | -2.58537600 | 7.24325900 |
| C | -1.20573200 | -1.82354200 | 8.86530300 |
| C | -0.65776900 | -1.13797200 | 9.96521300 |
| C | -2.29652400 | -2.69298000 | 9.08034600 |
| C | -1.19359300 | -1.30759600 | 11.25260300 |
| H | 0.19903600 | -0.46725500 | 9.79366300 |
| C | -2.83063400 | -2.86471500 | 10.36686900 |
| H | -2.72694600 | -3.24458000 | 8.22820200 |
| C | -2.28222200 | -2.16979400 | 11.45798300 |
| H | -0.75417100 | -0.75956800 | 12.10282900 |
| H | -3.68177600 | -3.54958600 | 10.51808300 |
| H | -2.70289100 | -2.30297600 | 12.46860000 |
| C | -7.71151000 | 0.60494200 | 0.17452700 |
| C | -6.95311600 | 1.11619700 | 1.25359200 |
| C | -8.53436800 | 1.51185200 | -0.53596100 |
| C | -7.01217100 | 2.47096500 | 1.60329700 |
| H | -6.29259700 | 0.43396500 | 1.81726300 |
| C | -8.58605000 | 2.86602400 | -0.17829000 |
| H | -9.17471400 | 1.14879100 | -1.35698900 |
| C | -7.82343600 | 3.37683100 | 0.89285700 |
| H | -6.41054400 | 2.85363700 | 2.44245000 |
| H | -9.26218100 | 3.53620200 | -0.73467800 |
| C | -7.81456600 | 4.87339900 | 1.26988900 |
| C | -6.76854300 | 5.63205600 | 0.43382800 |
| C | -9.21963900 | 5.49640200 | 1.13968700 |
| O | -7.39915200 | 5.01893100 | 2.64626600 |
| C | -6.63151400 | 5.43126700 | -0.95550900 |
| C | -5.94828100 | 6.60973400 | 1.02709900 |
| C | -9.44428200 | 6.78336400 | 0.61082500 |
| C | -10.33821300 | 4.81259900 | 1.66964300 |
| H | -8.19231500 | 4.82078800 | 3.19048600 |
| C | -5.73759100 | 6.18890200 | -1.71653800 |
| H | -7.23392300 | 4.66191700 | -1.46374600 |
| C | -5.04351400 | 7.37537600 | 0.27487900 |
| H | -6.03036600 | 6.77692400 | 2.11101500 |
| C | -10.71867200 | 7.36334300 | 0.60861900 |
| H | -8.59957400 | 7.34838800 | 0.18618400 |
| C | -11.61526700 | 5.38350200 | 1.68071200 |
| H | -10.21242000 | 3.79280800 | 2.07341600 |
| C | -4.92954200 | 7.19280300 | -1.12457000 |
| H | -5.65929000 | 6.00615500 | -2.80283800 |
| H | -4.44406300 | 8.14224500 | 0.78803700 |
| C | -11.83454600 | 6.67816300 | 1.14772800 |
| H | -10.85703500 | 8.37252900 | 0.18339100 |
| H | -12.46442600 | 4.81789900 | 2.10191200 |
| N | -0.05016700 | -7.82372600 | -0.95450800 |
| H | -7.06804000 | -1.29265800 | 0.49430800 |
| H | -6.04716400 | 0.03932600 | -2.41135600 |
| H | -2.55016600 | -1.16112900 | 6.81958200 |
| C | -0.16911700 | -9.03507600 | -0.17660300 |
| H | -1.14910700 | -8.99258900 | 0.36053500 |
| N | 0.84921700 | -9.12480600 | 0.87477800 |
| H | 0.61017200 | -9.86325000 | 1.55279000 |
| H | 1.76187300 | -9.36691500 | 0.45785300 |
| H | 0.46646700 | -7.91193400 | -1.83277900 |
| N | -4.10777700 | 7.96938500 | -1.94519500 |
| N | 0.33208000 | -0.55046600 | 7.37229200 |
| C | 1.59081000 | -0.62751400 | 6.76630200 |
| H | -0.13897900 | 0.34966700 | 7.23800800 |
| C | 2.39897600 | -1.78629900 | 6.84102600 |
| C | 2.10554200 | 0.49083600 | 6.06288300 |
| C | 3.63499600 | -1.84031300 | 6.17528400 |
| H | 2.07686800 | -2.65382300 | 7.43958100 |
| C | 3.34328200 | 0.43048000 | 5.41980500 |
| H | 1.50220200 | 1.41372000 | 6.00239700 |
| C | 4.12376300 | -0.74547200 | 5.44277300 |
| H | 4.23527100 | -2.76184800 | 6.24282400 |
| H | 3.71465200 | 1.30730100 | 4.86682300 |
| C | 5.42736900 | -0.78851400 | 4.63223100 |
| C | 5.09565800 | -0.91635000 | 3.12780800 |
| C | 6.36910400 | -1.90844800 | 5.10953500 |
| O | 6.07392800 | 0.47921400 | 4.89192000 |
| C | 5.85536700 | -0.22444500 | 2.16320600 |
| C | 4.04806500 | -1.72983200 | 2.65172300 |
| C | 6.99191900 | -1.77395400 | 6.37205300 |
| C | 6.66147400 | -3.06633200 | 4.36277000 |
| H | 7.01245100 | 0.37069100 | 4.62363700 |
| C | 5.60276500 | -0.36031100 | 0.79582700 |
| H | 6.65469700 | 0.46165200 | 2.48553300 |
| C | 3.77652500 | -1.87430400 | 1.28204700 |
| H | 3.40552900 | -2.26153400 | 3.37345000 |
| C | 7.86334800 | -2.74755100 | 6.86757900 |
| H | 6.78420900 | -0.87393500 | 6.97342200 |
| C | 7.53255400 | -4.05065000 | 4.85005100 |
| H | 6.20279300 | -3.20367900 | 3.37079500 |
| C | 4.56905100 | -1.20517900 | 0.31713200 |
| H | 6.21720800 | 0.20109300 | 0.07026300 |
| H | 2.92525400 | -2.50122200 | 0.97333700 |
| C | 8.15195500 | -3.91459500 | 6.11508000 |
| H | 8.33055700 | -2.61199600 | 7.85863100 |
| H | 7.74082700 | -4.94520000 | 4.23781000 |
| N | 4.41537800 | -1.36561100 | -1.05780000 |
| C | 3.23636600 | -1.91834800 | -1.70597100 |
| H | 4.89859200 | -0.65487300 | -1.61339900 |
| N | 2.04097000 | -1.07183600 | -1.67783200 |
| H | 2.94419900 | -2.82697100 | -1.14105000 |
| C | 2.02864400 | 0.25912300 | -2.07390200 |
| H | 1.33807200 | -1.32829900 | -0.96906000 |
| C | 2.98983100 | 0.81501900 | -2.96319200 |
| C | 1.01656300 | 1.13054600 | -1.59631700 |
| C | 2.95309700 | 2.17104100 | -3.30098800 |
| H | 3.76117000 | 0.17496900 | -3.41824100 |
| C | 0.98175600 | 2.48354100 | -1.95727800 |
| H | 0.24994100 | 0.72755100 | -0.91473700 |
| C | 1.96541800 | 3.04591900 | -2.79515300 |
| H | 3.71328800 | 2.57273700 | -3.99078900 |
| H | 0.17037000 | 3.11390700 | -1.56064100 |
| C | 1.98519200 | 4.50837300 | -3.26866100 |
| C | 3.35082500 | 5.17288500 | -3.02131000 |
| C | 0.90335400 | 5.39166700 | -2.60665600 |
| O | 1.80699100 | 4.50204100 | -4.70262800 |
| C | 4.14071900 | 4.87069100 | -1.89736300 |
| C | 3.82856700 | 6.16165400 | -3.90789300 |
| C | -0.01465200 | 6.11170100 | -3.38838900 |
| C | 0.83190200 | 5.56959900 | -1.20608700 |
| H | 1.05684100 | 3.89173700 | -4.87636500 |
| C | 5.35695700 | 5.52602100 | -1.64968100 |
| H | 3.81645000 | 4.08052200 | -1.20001100 |
| C | 5.03495700 | 6.82519100 | -3.67349500 |
| H | 3.23596100 | 6.40952000 | -4.80170700 |
| C | -0.97223300 | 6.96597200 | -2.81716900 |
| H | 0.03215600 | 6.00905800 | -4.48362100 |
| C | -0.11854300 | 6.41041800 | -0.62310000 |
| H | 1.54657500 | 5.04662700 | -0.54967800 |
| C | 5.82793700 | 6.53143400 | -2.53222500 |
| H | 5.95983500 | 5.22311200 | -0.77949000 |
| H | 5.38234200 | 7.59547400 | -4.38439700 |
| C | -1.04093100 | 7.14258200 | -1.41568400 |
| H | -1.66202000 | 7.50012800 | -3.48968600 |
| H | -0.13954800 | 6.53126500 | 0.47403400 |
| N | 7.02709100 | 7.19746600 | -2.32891900 |
| N | -1.92049400 | 8.02853600 | -0.79280600 |
| C | 7.69247100 | 7.32462500 | -1.03080900 |
| H | 7.21478100 | 8.00695000 | -2.92615300 |
| C | -2.96216000 | 8.75841100 | -1.48578600 |
| H | -2.08236300 | 7.86031700 | 0.20310800 |
| H | 6.90482100 | 7.19030100 | -0.24331600 |
| N | 8.28057700 | 8.65740200 | -0.94298800 |
| H | -2.50121500 | 9.13616200 | -2.42650900 |
| C | -3.38061000 | 10.00560000 | -0.69186400 |
| H | 7.75209600 | 9.28326200 | -0.32408400 |
| H | 9.26235600 | 8.61245200 | -0.64347300 |
| C | -2.48075700 | 10.63634000 | 0.19066600 |
| C | -4.64787800 | 10.58718400 | -0.89724500 |
| C | -2.84487300 | 11.81541900 | 0.86241300 |
| H | -1.47987800 | 10.20178700 | 0.34196300 |
| C | -5.01066000 | 11.76560500 | -0.22807300 |
| H | -5.35171700 | 10.09616800 | -1.58770100 |
| C | -4.11159900 | 12.38394100 | 0.65704700 |
| H | -2.12755400 | 12.29436100 | 1.54978600 |
| H | -6.00894400 | 12.20301100 | -0.39688500 |
| H | -4.39846500 | 13.30844300 | 1.18530700 |
| H | -3.98120100 | 7.57278900 | -2.87892400 |
| N | -13.08897200 | 7.27024900 | 1.19461500 |
| H | -13.89300600 | 6.64977400 | 1.31243000 |
| H | -13.25740800 | 8.05346700 | 0.55934600 |
| N | 8.97570000 | -4.90971600 | 6.62526100 |
| H | 9.37795200 | -5.56729900 | 5.95340400 |
| H | 9.61592700 | -4.64134700 | 7.37601500 |
| C | 8.72870900 | 6.23897700 | -0.79694600 |
| C | 9.64223500 | 6.16573400 | 0.32185700 |
| C | 8.92734300 | 5.05624200 | -1.59660500 |
| Fe | 10.66801000 | 6.11983600 | -1.44624300 |
| C | 10.39903300 | 4.94753000 | 0.20628000 |
| H | 9.73945800 | 6.90758500 | 1.12788500 |
| C | 9.95584000 | 4.26225000 | -0.97992600 |
| H | 8.38593100 | 4.82861700 | -2.52410600 |
| C | 10.99723400 | 7.68620700 | -2.73094900 |
| H | 11.18498800 | 4.60793700 | 0.89567300 |
| H | 10.34848800 | 3.30809600 | -1.35873300 |
| C | 11.83634600 | 7.80180500 | -1.56564400 |
| C | 11.29489500 | 6.43049700 | -3.36779200 |
| H | 10.24435200 | 8.41698600 | -3.05646200 |
| C | 12.65206500 | 6.61804400 | -1.48278900 |
| H | 11.86166800 | 8.64914400 | -0.86505700 |
| C | 12.31877400 | 5.77166000 | -2.59888300 |
| H | 10.81230800 | 6.03706400 | -4.27332200 |
| H | 13.39438000 | 6.39751200 | -0.70287700 |
| H | 12.75786100 | 4.78801800 | -2.81722100 |
| C | 3.60795800 | -2.34556400 | -3.12490700 |
| C | 2.66935400 | -2.71383900 | -4.15402900 |
| C | 4.93277200 | -2.42938800 | -3.69678700 |
| Fe | 4.00048400 | -4.23849700 | -3.82485500 |
| C | 3.40494000 | -3.01427800 | -5.35240800 |
| H | 1.57829600 | -2.74434100 | -4.03136900 |
| C | 4.80494900 | -2.83624600 | -5.07168500 |
| H | 5.87316500 | -2.24376800 | -3.16124800 |
| C | 4.76180800 | -5.47118400 | -2.37701000 |
| H | 2.97331600 | -3.33585800 | -6.31056600 |
| H | 5.63334000 | -3.00068700 | -5.77504900 |
| C | 3.33151100 | -5.60968500 | -2.45868900 |
| C | 5.31659700 | -5.79871800 | -3.66451700 |
| H | 5.32787600 | -5.14664500 | -1.49252300 |
| C | 3.00258400 | -6.01977500 | -3.80094200 |
| H | 2.61251400 | -5.41136400 | -1.65004500 |
| C | 4.22993700 | -6.13960300 | -4.54471500 |
| H | 6.38173300 | -5.77428200 | -3.93409400 |
| H | 1.99084100 | -6.18672400 | -4.19888100 |
| H | 4.31959200 | -6.42380500 | -5.60255400 |
| C | -7.87510500 | -2.64453400 | -1.70201300 |
| C | -7.53909300 | -3.48137800 | -2.82706400 |
| C | -8.65454500 | -3.43869800 | -0.78235900 |
| Fe | -9.57091100 | -3.36564800 | -2.60786700 |
| C | -8.09793200 | -4.78647700 | -2.59795800 |
| H | -6.94546200 | -3.16428700 | -3.69471700 |
| C | -8.78776300 | -4.76089700 | -1.33493000 |
| H | -9.07961800 | -3.07417100 | 0.16182200 |
| C | -10.35526600 | -1.99054100 | -3.90511600 |
| H | -8.02430800 | -5.64718000 | -3.27734700 |
| H | -9.33671200 | -5.59751000 | -0.88035700 |
| C | -10.50735600 | -3.32381800 | -4.42713400 |
| C | -11.03012300 | -1.93022800 | -2.63504000 |
| H | -9.80774400 | -1.16878400 | -4.38888600 |
| C | -11.27799400 | -4.08633500 | -3.48042900 |
| H | -10.09342000 | -3.69761600 | -5.37408500 |
| C | -11.59936300 | -3.22619700 | -2.37213600 |
| H | -11.08526000 | -1.05650500 | -1.96955500 |
| H | -11.55409800 | -5.14576000 | -3.57710600 |
| H | -12.16459900 | -3.51217500 | -1.47411300 |
| C | -0.23142900 | -10.22357600 | -1.12288100 |
| C | 0.20217600 | -11.56627600 | -0.81671300 |
| C | -0.81977900 | -10.25243600 | -2.44234500 |

```
Fe        1.12906500  -10.89122400  -2.50549100
C        -0.11045100  -12.41127000  -1.93783900
H         0.70044100  -11.88240400   0.11014000
C        -0.74026100  -11.59744100  -2.94413200
H        -1.25607600   -9.39034800  -2.96493400
C         3.14210000  -10.87716100  -2.11717300
H         0.11024700  -13.48496100  -2.01786300
H        -1.08665000  -11.94105400  -3.92915800
C         2.82104700  -11.84840500  -3.13010500
C         2.72191800   -9.58396700  -2.58964000
H         3.62927100  -11.08803000  -1.15458600
C         2.19639000  -11.15624800  -4.22735000
H         3.00934800  -12.92961600  -3.07129000
C         2.13338800   -9.75827300  -3.89180400
H         2.85545200   -8.62518800  -2.06791600
H         1.82452600  -11.61602400  -5.15372000
H         1.71816100   -8.96002800  -4.52277200
```

| | |
|---|---|
| Zero-point correction = | 2.202510 (Hartree/Particle) |
| Thermal correction to Energy = | 2.341599 |
| Thermal correction to Enthalpy = | 2.342544 |
| Thermal correction to Gibbs Free Energy = | 1.996116 |
| Sum of electronic and zero-point Energies = | -11285.954002 |
| Sum of electronic and thermal Energies = | -11285.814913 |
| Sum of electronic and thermal Enthalpies = | -11285.813968 |
| Sum of electronic and thermal Free Energies = | -11286.160396 |
| HF= -11288.1565121 | |

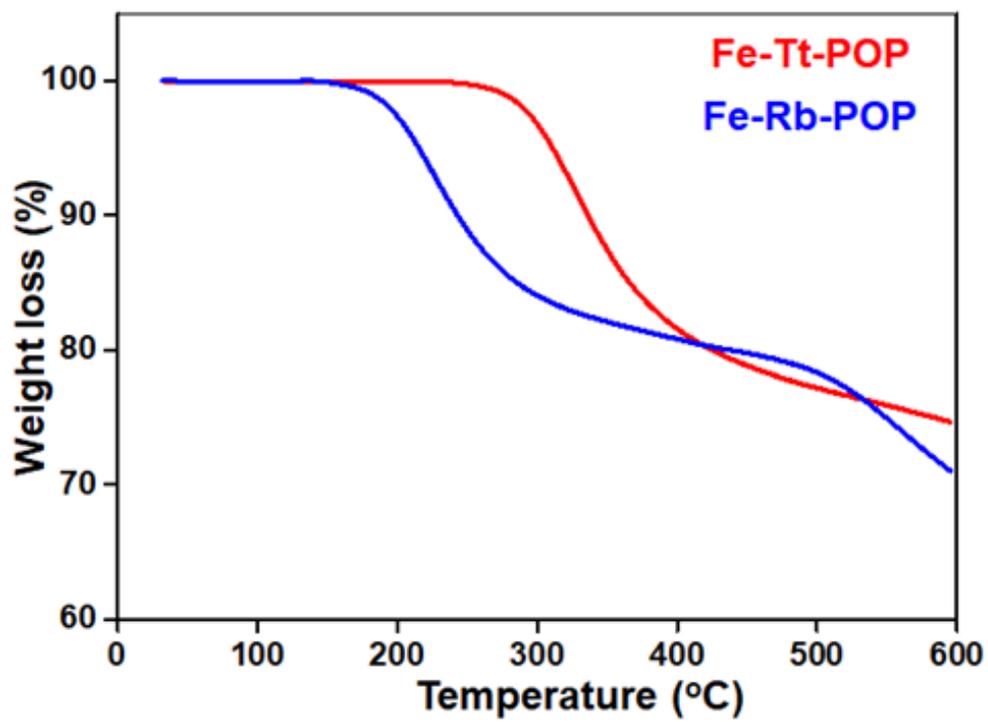

**Figure S2:** Thermo Gravimetric Analysis (TGA) of **Fe-Rb-POP** & **Fe-Tt-POP**.

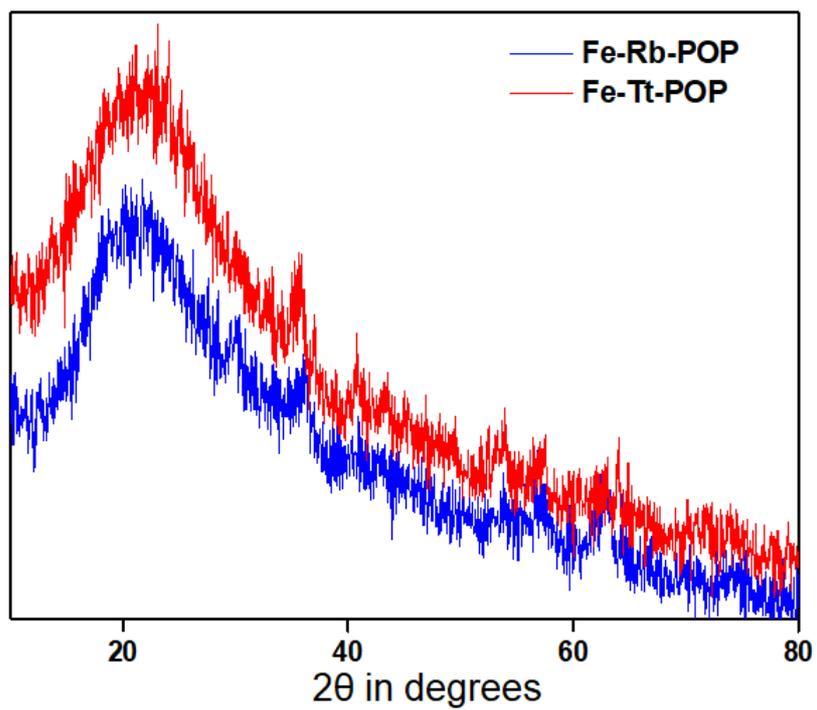

**Figure S3:** Wide angle PXRD pattern of the respective **Fe-POPs**.

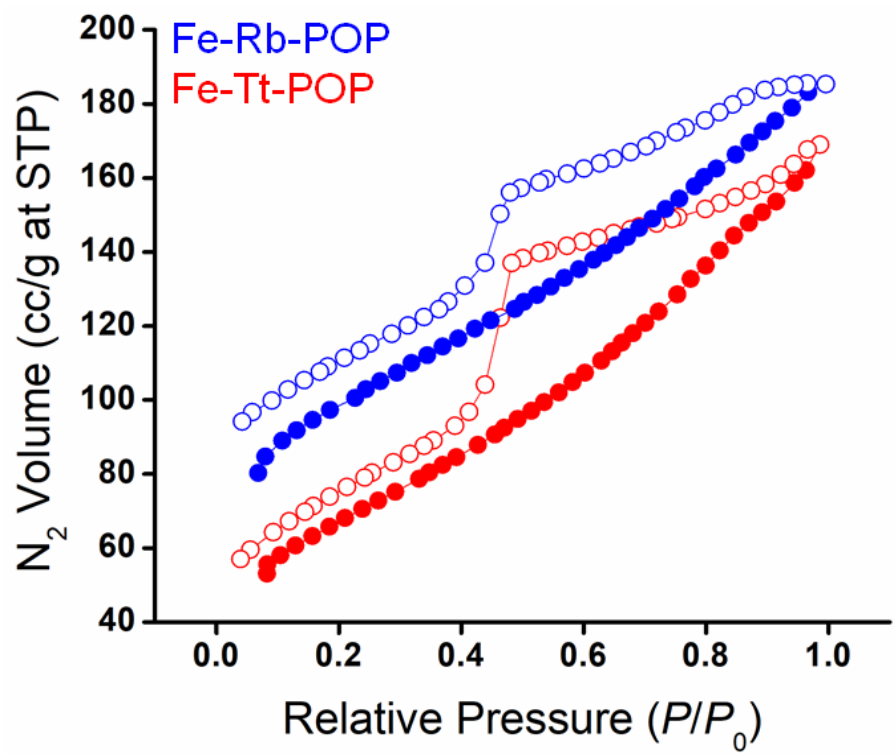

**Figure S4:** N$_2$-adsorption/desorption isotherms as measured at 77 K of the respective **Fe-POPs**.

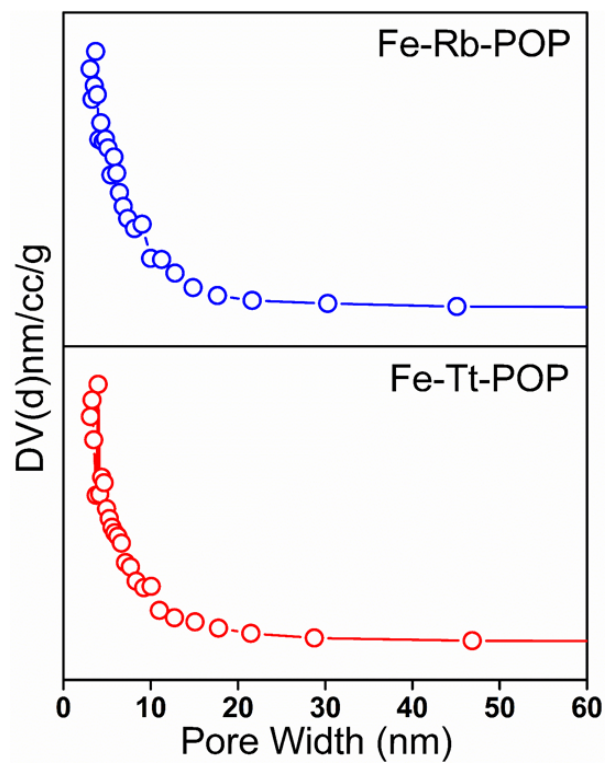

**Figure S5:** Pore-size distributions of the respective **Fe-POPs** as calculated by NLDFT method.

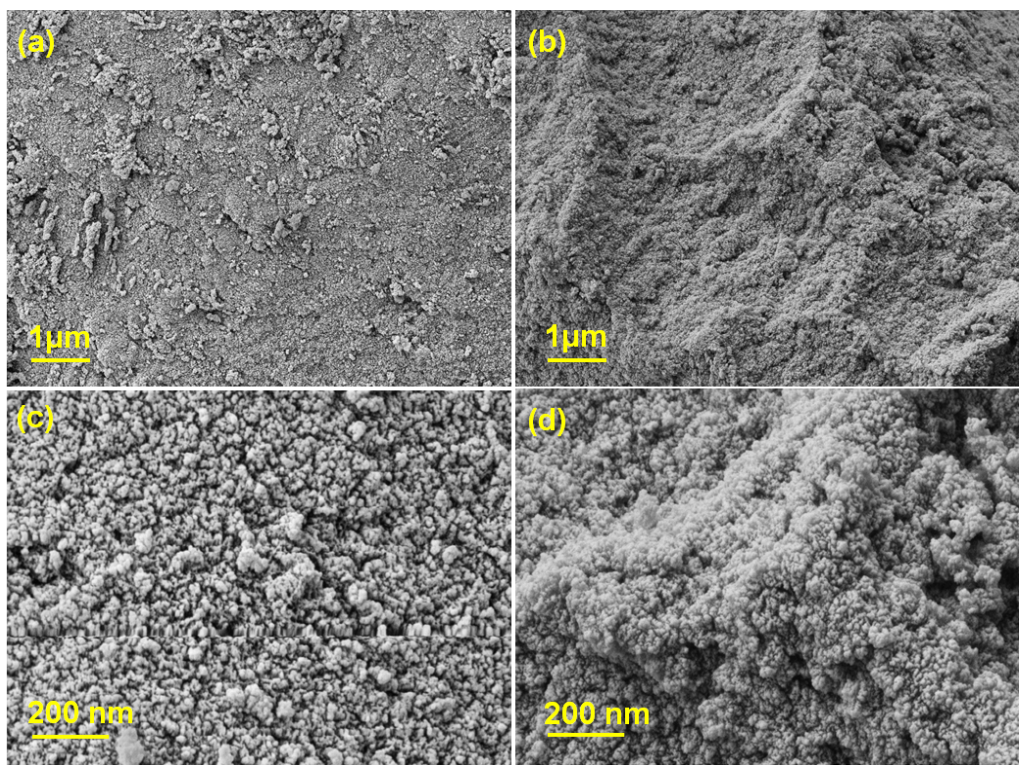

**Figure S6:** FE-SEM images (a & b) of **Fe-Rb-POP** & (c & d) of **Fe-Tt-POP**, respectively.

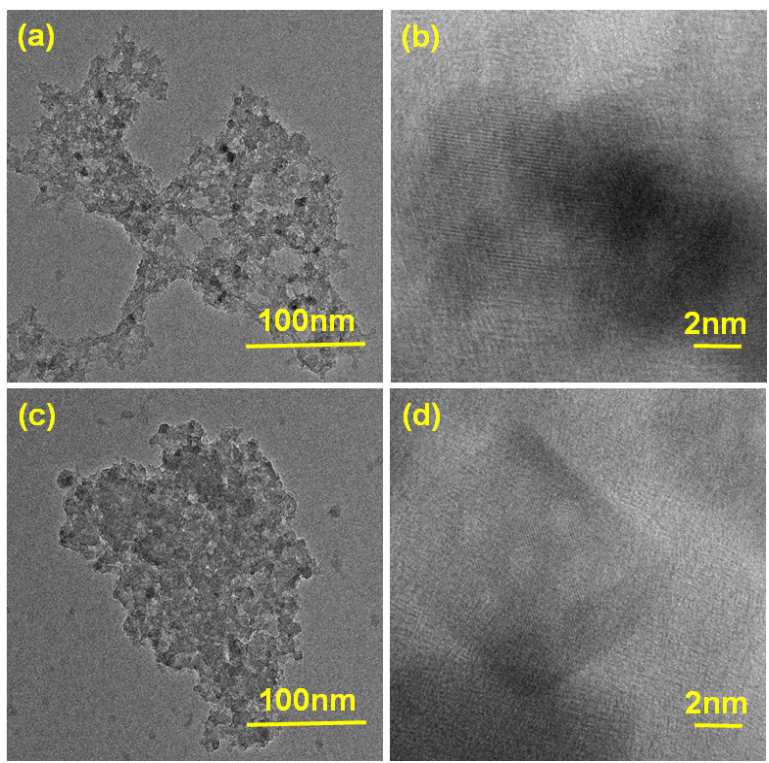

**Figure S7:** TEM images (a & b) of **Fe-Rb-POP** & (c & d) of **Fe-Tt-POP**, respectively.

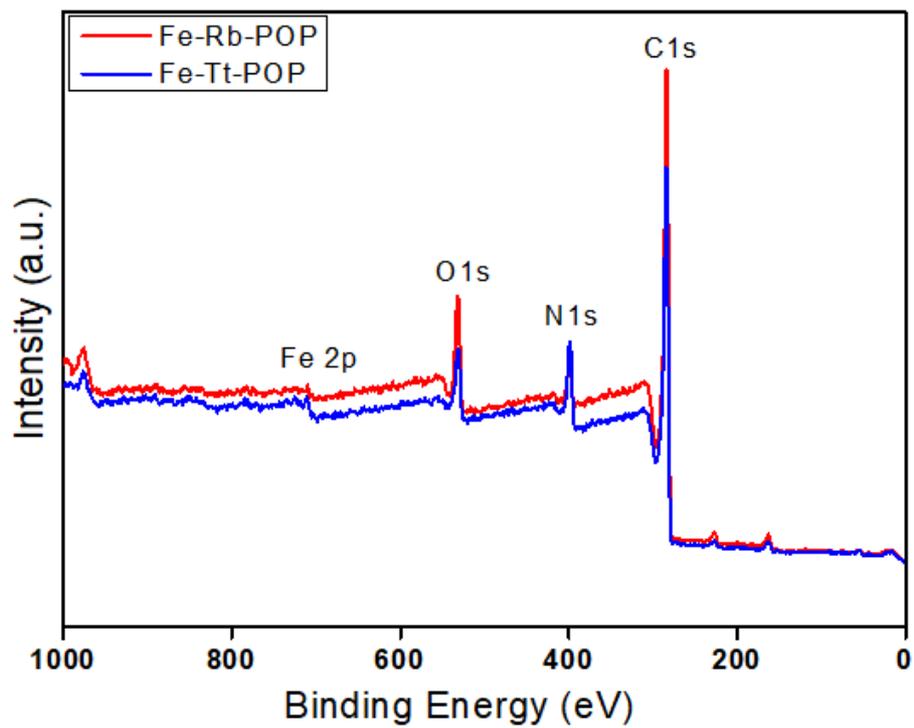

**Figure S8:** XPS survey spectra of the respective **Fe-POPs**.

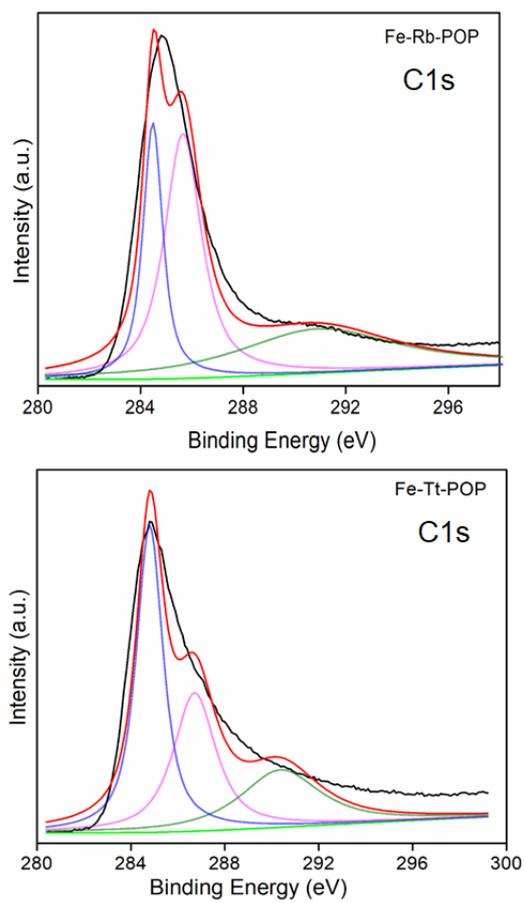

**Figure S9:** XP-spectra at C-1s core region of the respective **Fe-POPs**.

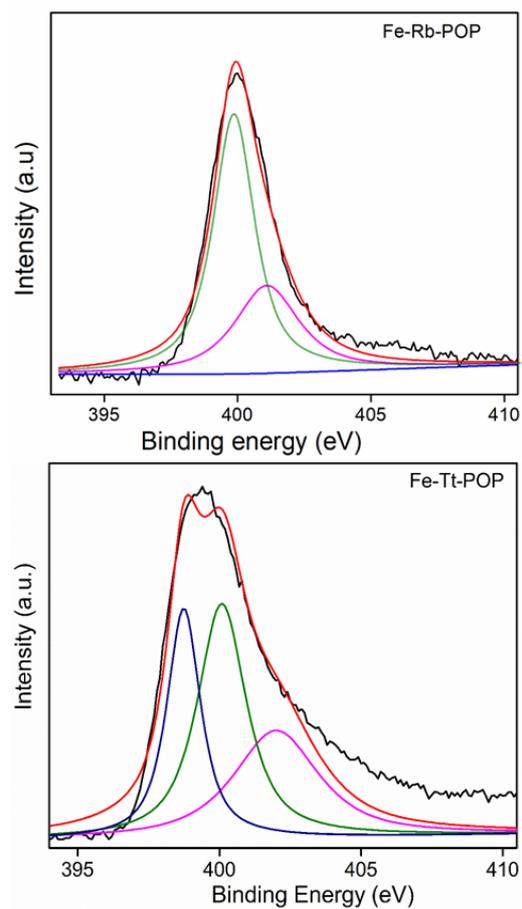

**Figure S10:** XP-spectra at N-1s core region of the respective **Fe-POPs**.

**Table S3:** The table of product distribution under the influence of different reaction conditions such as without catalyst, without $H_2O_2$, under $N_2$ atmosphere, and at room temperature.

| Entry | Catalyst | 100 $^0$C | Variation | Time (hour) | MPS Yield (%) |
|---|---|---|---|---|---|
| 1[a] | **Fe-Tt-POP** | + | - | 13 | 94% |
| 2[b] | **Fe-Rb-POP** | + | - | 13 | 43% |
| 3[c] | **Without catalyst** | + | - | 13 | trace |
| 4[d] | **Fe-Tt-POP** | + | No $H_2O_2$ | 13 | trace |
| 5[d] | **Fe-Rb-POP** | + | No $H_2O_2$ | 13 | trace |
| 6[e] | **Fe-Tt-POP** | + | $N_2$ | 13 | trace |
| 7[e] | **Fe-Rb-POP** | + | $N_2$ | 13 | trace |
| 8[f] | **Fe-Tt-POP** | - | Room Temp. | 13 | trace |
| 9[f] | **Fe-Rb-POP** | - | Room Temp. | 13 | trace |

**Reaction conditions:** TA (0.5 mmol), Fe-POPs (50mg), acetonitrile (20 mL), 100°C, $H_2O_2$ (0.25 ml). For **Fe-Tt-POP**[a] & **Fe-Rb-POP**[b] Isolated yield, without using any catalyst[c], without $H_2O_2$ [d], in $N_2$ atmosphere [e], in room temperature[f].

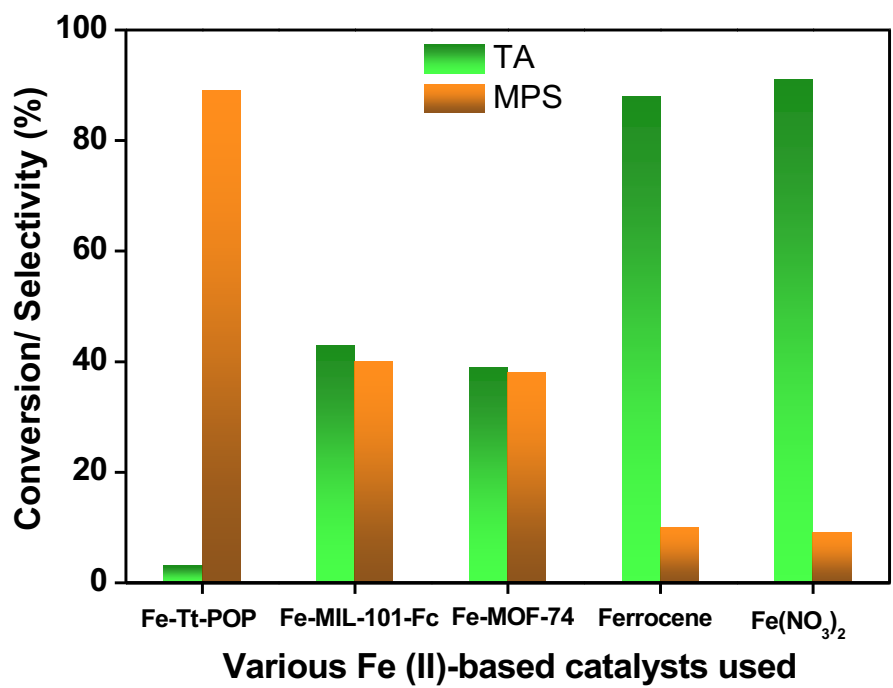

**Figure S11:** Catalytic property tests with various Fe (II)-based catalysts under optimized reaction conditions.

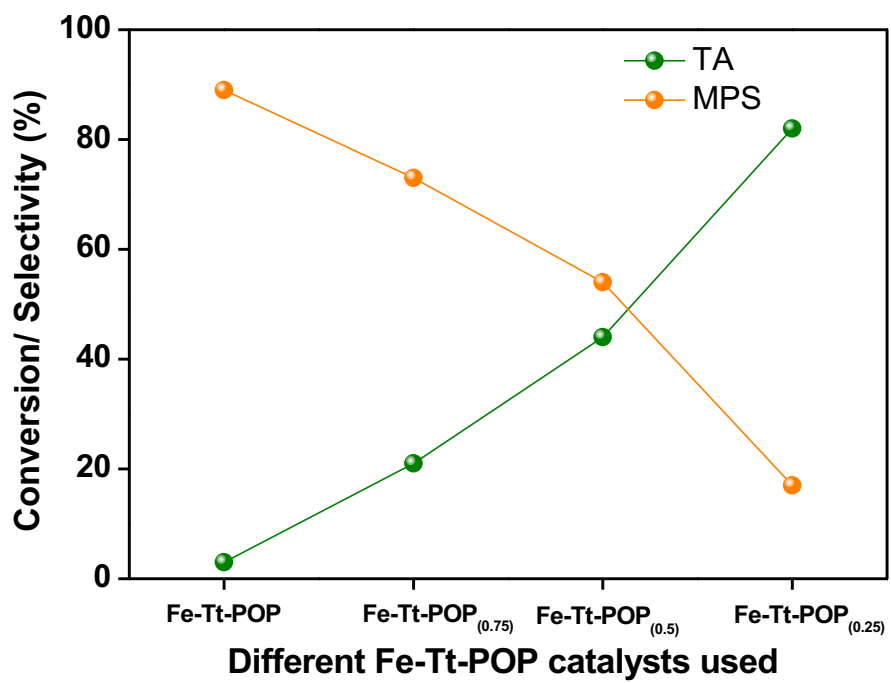

**Figure S12:** Catalytic property tests with **Fe-Tt-POP** catalyst with various Fe contents under optimized reaction conditions.

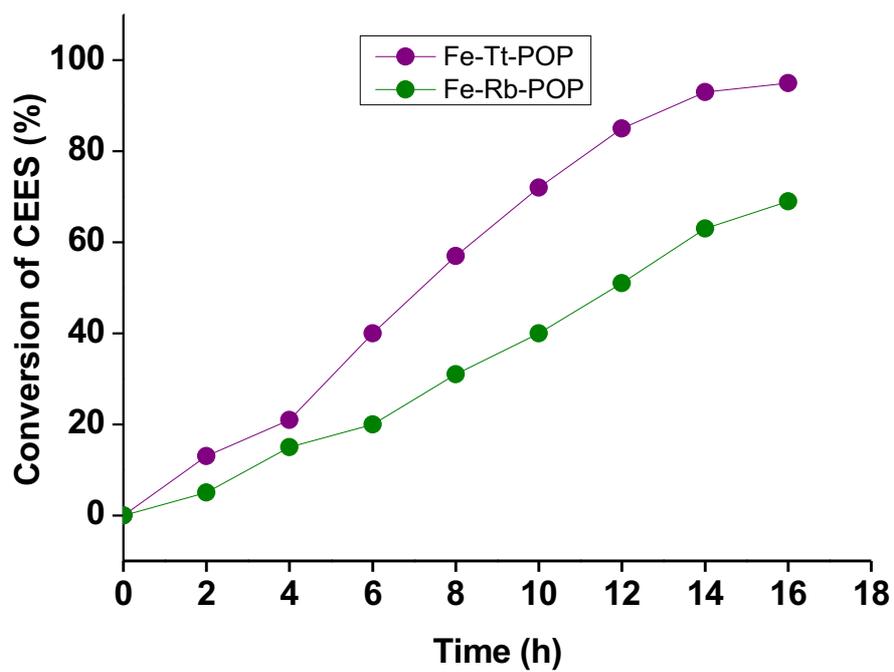

**Figure S13:** Time profile for the oxidative decontamination of CEES using **Fe-Tt-POP** & **Fe-Rb-POP** catalyst under optimized reaction conditions.

**Table S4:** Calculated ΔG values of sulfoxide-O and sulfoxide-S pathway for Fe-Tt-POP and Fe-Rb-POP, respectively

| Free Energy (ΔG) (eV) | | |
|---|---|---|
| **Reaction Pathway** | **Fe-Tt-POP (planar)** | **Fe-Rb-POP (non-planar)** |
| Sulfoxide-O | -0.061 | -0.097 |
| Sulfoxide-S | -0.017 | -0.055 |

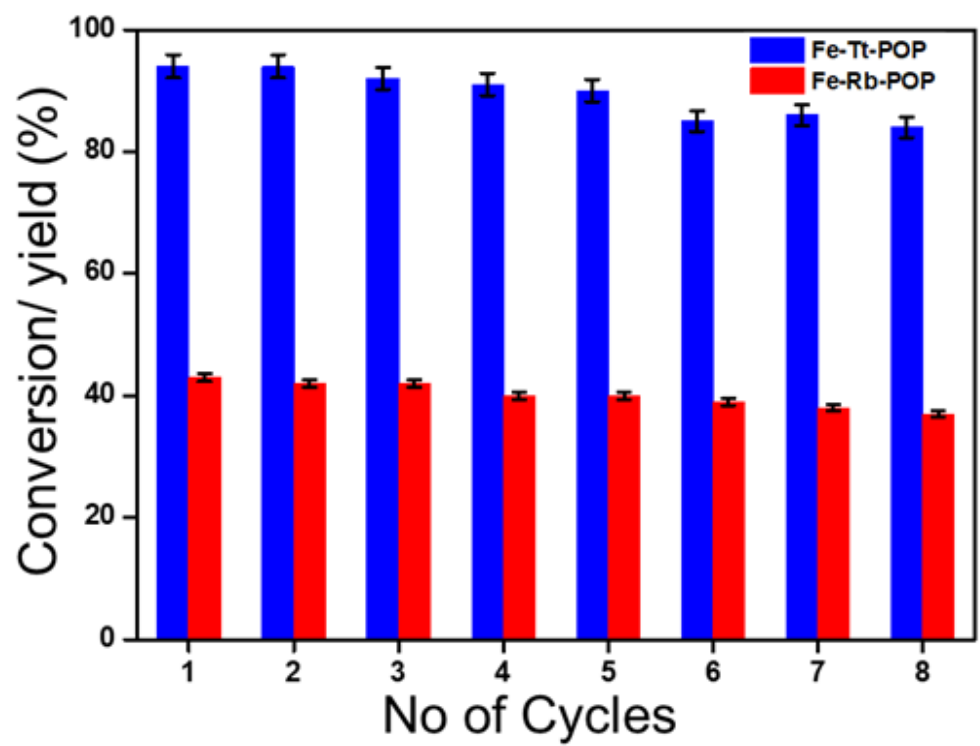

**Figure S14:** Reusability test of **Fe-Tt-POP** and **Fe-Rb-POP**. Reaction conditions: TA (2.5 mmol), $H_2O_2$ (1.25 ml), $CH_3CN$ (100 ml), catalyst (250 mg), temperature (100 °C), and time (13 h).

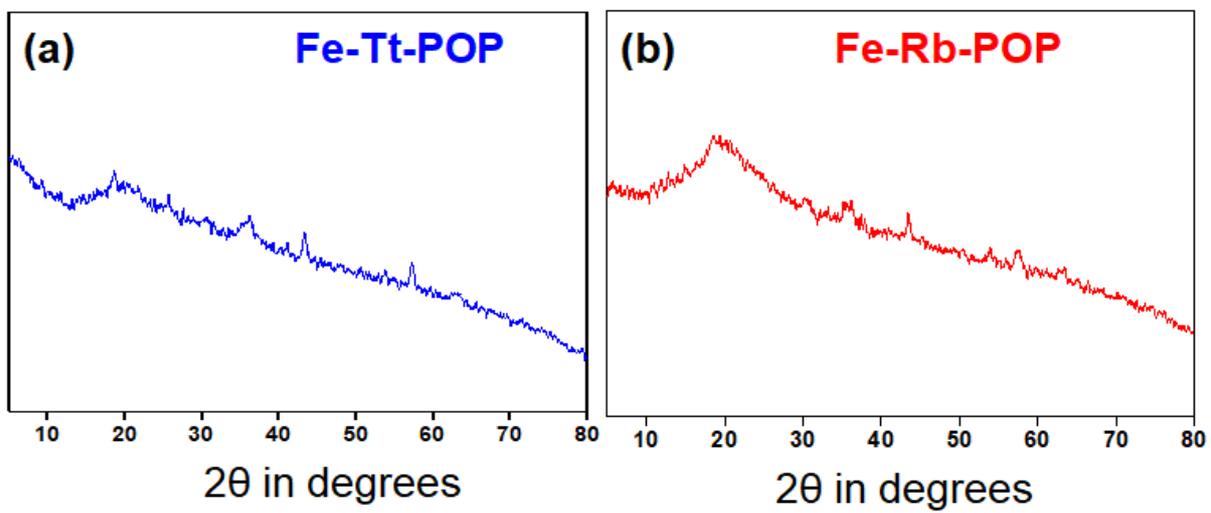

**Figure S15:** Reused catalyst PXRD analysis of **Fe-Tt-POP** (a) and **Fe-Rb-POP** (b), respectively.

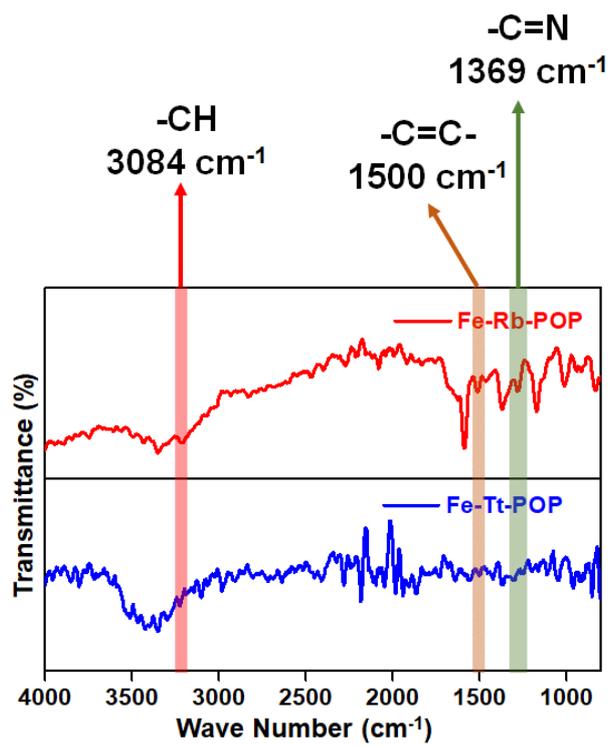

**Figure S16:** FT-IR analysis of reused catalyst **Fe-Tt-POP** & **Fe-Rb-POP**, respectively.

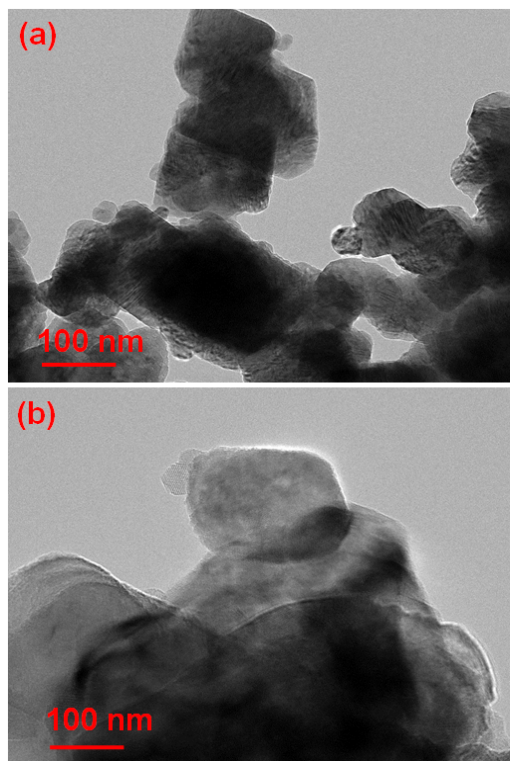

**Figure S17:** TEM images of reused catalysts **Fe-Tt-POP** (a) and **Fe-Rb-POP** (b), respectively after 8th consecutive catalytic run.

**Table S5:** Comparison study in thioanisole conversion to methyl phenyl sulfoxide with other previous reported catalysts under various reaction conditions.

| Entry | Previously Reported Catalyst | Conditions | Conversion | Ref. |
|---|---|---|---|---|
| 1 | 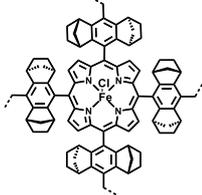 | alkene (400 μmol), PhIO (200 μmol) and catalyst (1 μmol) CH$_2$Cl$_2$ (1 ml) 5 h, Substrate 0.5 mmol | 43% | 9 |
| 2 | 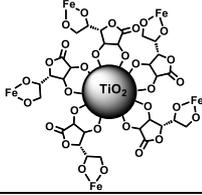 | sulfide/TBHP/Cat, 1000: 2000:3 and the reaction were run in solvent free condition at 60° C. | 75% | 10 |
| 3 | 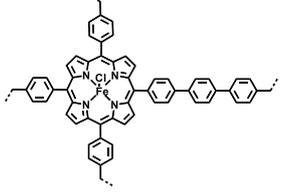 | FeP -CMP (5.86 × 10-7 mol Fe), toluene (3 mL), room temperature (rt), O$_2$ (1 atm), Fe/substrate/IBA 1/1000/1000 (molar ratio) | 99% | 11 |
| 4 | 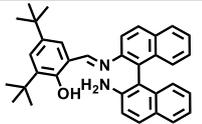<br>Ligand | H$_2$O$_2$ (1.2 equiv.), Ligand (4.0 mol%), FeCl$_3$X 6H$_2$O (2.0 mol%), THF, rt, 24 h | 74% | 12 |
| 5 | 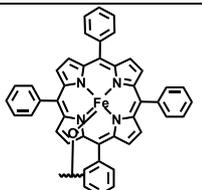 | 1000:3000:1 for sulfide/TBAOX/catalyst, rt, air, H$_2$O | 92% | 13 |
| 6 | 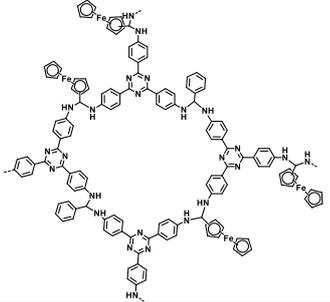 | **TA (0.5 mmol), H$_2$O$_2$ (0.25 ml), CH$_3$CN (10 ml), catalyst (50 mg), temperature (100 °C), and time (13 h)** | 94% | **Our Work** |